\pdfoutput=1
\documentclass{aa}
\usepackage{natbib}
\bibpunct{(}{)}{;}{a}{}{,} % to follow the A&A style
\usepackage{graphicx}
\usepackage{txfonts}
\usepackage[colorlinks=true,linkcolor=blue,allcolors=blue]{hyperref}%
\usepackage{subfigure}

\makeatletter
\renewcommand*\aa@pageof{, page \thepage{} of \pageref*{LastPage}}
\makeatother

\begin{document}

   \title{Single site observations of \textit{TESS} single transit detections}

   \author{Benjamin F. Cooke\inst{1,2}
   \and Don Pollacco\inst{1,2}
   \and Richard West\inst{1,2}
   \and James McCormac\inst{1,2}
   \and Peter J. Wheatley\inst{1,2}
          }

   \institute{Department of Physics, University of Warwick, Gibbet Hill Road, Coventry CV4 7AL, UK\label{inst1}
   \and Centre for Exoplanets and Habitability, University of Warwick, Gibbet Hill Road, Coventry CV4 7AL, UK\label{inst2}
             }

   %\date{Received September 15, 1996; accepted March 16, 1997}

\abstract
{\textit{TESS} has been successfully launched and has begin data acquisition. To expedite the science that may be performed with the resulting data it is necessary to gain a good understanding of planetary yields. Given the observing strategy employed by \textit{TESS} the probability of detecting single transits in long period systems is increased. These systems require careful consideration.}
{To simulate the number of \textit{TESS} transit detections during its 2 year mission with a particular emphasis on single transits. Additionally, to determine the feasibility of ground-based follow-up observations from a single site.}
{A distribution of planets is simulated around the $\sim$ 4 million stars in the \textit{TESS} Candidate Target List. These planets are tested for detectable transits and characterised. Based on simulated parameters the single transit detections are further analysed to determine which are amenable to ground-based follow-up.}
{\textit{TESS} will discover an approximate lower bound of 4700 planets with around 460 being single transits. A large fraction of these will be observable from a single ground-based site. This paper finds that, in a single year, approximately 1000 transit events of around 320 unique \textit{TESS} single transit detections are theoretically observable.}
{As we consider longer period exoplanets the need for exploring single transit detections increases. For periods $\gtrsim45$ days the number of single transit detections outnumber multitransits by a factor of 3 (82$\pm$18 and 25$\pm$7 respectively) a factor which only grows as longer period detections are considered. Therefore, based on the results of this paper, it is worth expending the extra effort required to follow-up these more challenging, but potentially very rewarding, discoveries. Additionally, we conclude that a large fraction of these targets can be theoretically observed from just a single ground-based site. However, further work is required to determine whether these follow-up efforts are feasible when accounting for target specific criteria.}
% 5 {} token are mandatory


   \maketitle
%
%________________________________________________________________

\section{Introduction}
\label{sec:Introduction}

As of 28 June 2018 there are 3735 confirmed exoplanet discoveries \citep[NASA Exoplanet Archive\footnote{\href{https://exoplanetarchive.ipac.caltech.edu/index.html}{https://exoplanetarchive.ipac.caltech.edu/index.html}},][]{Akeson2013} with 2934 having been discovered via the transit method. The vast majority of these detections come from a small number of efficacious surveys including \textit{Kepler} \citep[][2327 detections]{Borucki2010}, \textit{K2} \citep[][294 detections]{Howell2014}, the Super Wide Angle Search for Planets \citep[\textit{SuperWASP},][125 detections]{Pollacco2006}, the Hungarian Automated Telescope survey \citep[\textit{HAT},][101 detections]{Bakos2004}, the Convection, Rotation and planetary Transits satellite \citep[\textit{CoRoT},][30 detections]{Baglin2003} and the Kilodegree Extremely Little Telescope survey \citep[\textit{KELT},][19 detections]{Siverd2012}. The Transiting Exoplanet Survey Satellite \citep[\textit{TESS},][]{Ricker2015} is expected to discover on the order of $10^3$ exoplanets \citep{Sullivan2015} which will make it the most prolific exoplanet discovery mission ever.

\textit{TESS} launched on 18 April 2018 and will search for transiting planets around bright, nearby stars \citep{Ricker2015}. \textit{TESS} will be unrivalled in its ability to find hundreds of sub-Neptune radius planets around bright stars that provide amenable targets for follow-up and atmospheric characterisation \citep{Crouzet2017}. Thanks to its large viewing area (>85\% of the sky will be observed to some degree) current simulations predict \textit{TESS} will discover thousands of planets orbiting stars with $m_V\leq12$ \citep{Sullivan2015,Bouma2017,Ballard2018,Barclay2018,Villanueva2018}. Subsequent follow-up observations will enable the determination of planet mass via radial velocity measurements \citep{Cloutier2018} and the characterisation of planet atmosphere via transmission spectroscopy \citep{Kempton2018}.

The observing strategy employed by \textit{TESS}, combined with the fact that some areas of the sky will be observed for long baselines means that the chance of detecting the transits of long period planets is increased compared to shorter duration missions. For example, each \textit{K2} field is observed for $\sim75$ days \citep{Howell2014} whereas some regions of \textit{TESS} fields will reach baselines of >300 days \citep{Ricker2015}. However, long period orbits then mean that it is likely that only a single transit event will be observed. Traditionally, transit detections are confirmed using many repeated transits folded along the period of the orbit leading to improved signal detection and signal to noise ratio, S/N. Obviously this is not possible with single transits and so these need to be treated differently to avoid losing potential discoveries. Using \textit{Kepler} and \textit{K2} as a guide it has been shown that observations of just a single transit can still lead to accurate orbital parameters on the $10-20\%$ level at 30-minute cadence \citep{Yee2008,Osborn2015} meaning they are still worth the additional effort required to confirm. Follow-up however, will need to be more rigorous, since the ephemerides are less constrained than for a system with multiple observed transits, and will require significant follow-up effort from high-precision ground-based facilities, such as the Next Generation Transit Survey (\textit{NGTS}, \citealt{Wheatley2017}) and space-based observatories such as the James Webb Space Telescope (\textit{JWST}, \citealt{Gardner2006}).

To this end a detailed simulation of the number of \textit{TESS} single transits must be carried out and then extended to determine how many of the events may be observed from the ground to confirm the periodic nature of the system. We therefore create a simulation of \textit{TESS} detections of single transits and then determine their follow-up observability from a single ground-based site.

We set out the paper in the following way. Section \ref{sec:Stellar population} discusses the stellar population used in the project with Sect. \ref{sec:Planet population} describing the generation of a planetary population. Next, Sect. \ref{sec:Detectability} determines the detectability of these simulated planets by \textit{TESS} and Sect. \ref{sec:Single site follow-up} details their potential for single site follow-up observations. Section \ref{sec:Analysis and Results} displays the simulation results and finally, Sect. \ref{sec:Discussion and conclusions} outlines some points of discussion and the conclusions.

\section{Stellar population}
\label{sec:Stellar population}

\begin{figure}[ht]
	\includegraphics[width=\columnwidth]{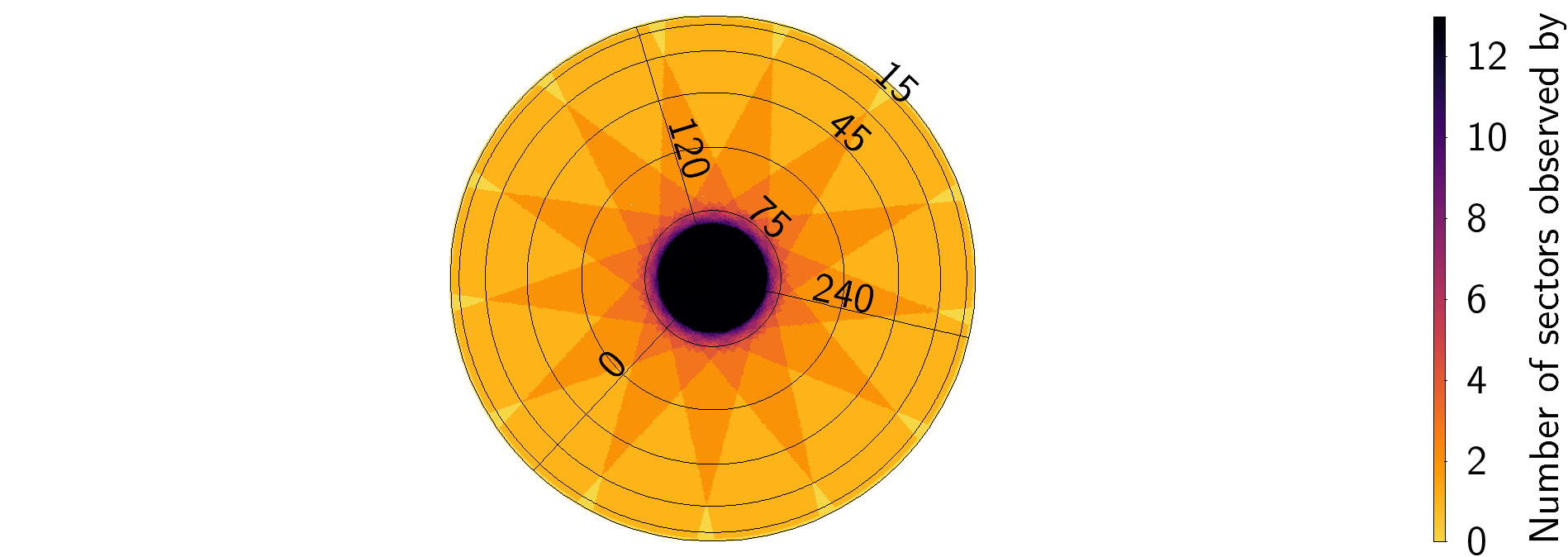}
    \caption{Plot of all TIC CTL stars in the northern ecliptic hemisphere coloured by number of observing regions. The north ecliptic is shown but the pattern is mirrored on the southern ecliptic hemisphere.}
    \label{fig:TESS sectors}
\end{figure}

Our simulation is built using the 3.8 million stars in the TESS Input Catalogue (TIC) Candidate Target List (CTL) available from the Mikulski Archive for Space Telescopes (MAST\footnote{\href{https://archive.stsci.edu/tess/bulk_downloads.html}{https://archive.stsci.edu/tess/bulk\_downloads.html}}). The key parameters within the CTL are stellar properties including \textit{TESS} magnitude ($m_{TESS}$), mass ($M_{\star}$), radius ($R_{\star}$), effective temperature ($T_{eff})$ and luminosity ($L_{\star}$) as well as positional parameters such as R.A. \& Dec., latitude \& longitude (ecliptic) and contamination ratio (for a full description of the CTL and its parameters see \citealt{Stassun2017}). Additionally, it is necessary to know for how long each star (and therefore each potential planet) will be observed for. To determine this baseline value we simulate 13 sectors (per ecliptic hemisphere) on the sky, each one a $24^\circ\times96^\circ$ rectangular region \citep{Ricker2015}. Each CTL star was then checked using ecliptic coordinates and it was determined how many sectors would cover this region and which sectors (numbered from 1-13) they would be. It was assumed that the 13 sectors were evenly distributed around the sky and reflected exactly about the ecliptic equator. Additionally, since the initial spacecraft pointing is not yet known \citep{Bouma2017}, it was decided that the first sector would be centred on the line of longitude = 0$^{\circ}$. Figure \ref{fig:TESS sectors} shows a plot of all CTL stars coloured by the number of regions for which they will be observed. The observing baseline for each target is then taken to be 27.4 days multiplied by the number of sectors which observe that target \citep{Ricker2015}. As an additional caveat, \textit{TESS} will spend $\sim$6.5 hrs per 13.7 day orbit transferring data during which time data acquisition will halt. This time has been included in these simulations as non-observing time. We can then simulate a synthetic population of planets around these stars and the systems can be checked for detections.

\begin{figure}[ht]
	\includegraphics[width=\columnwidth]{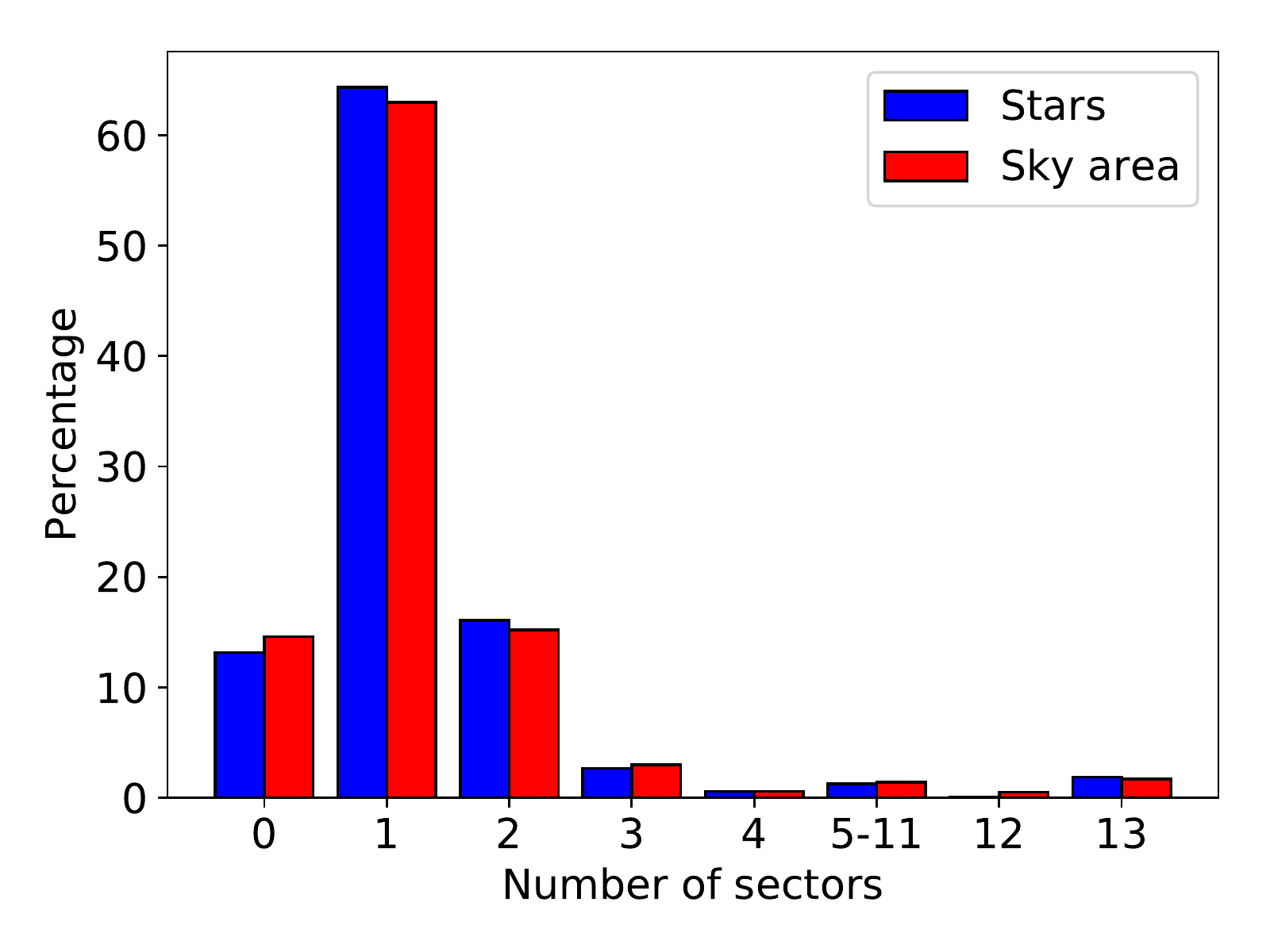}
    \caption{Plot of \textit{TESS} sky coverage showing area of sky observed by combinations of regions. Sky areas observed by more than 4 and less than 12 regions are combined as individually these regions are very small. The plot shows sky area in red and number of stars in blue, both given as a percentage of the total.}
    \label{fig:sector analysis}
\end{figure}

Figure \ref{fig:sector analysis} shows the percentage of sky covered by combinations of sectors; from zero along the ecliptic equator, up to 13 in the \textit{TESS} continuous viewing zone (CVZ). This is then shown alongside the percentage of CTL stars observed by the same number of sectors. Discrepancies between the two sets of results is due to the non-uniformity of CTL stars when distributed across the celestial sphere (for example, stellar density is increased in the galactic plane).

As an additional step we estimate the spectral type of each star based on their $T_{eff}$ values. The chosen $T_{eff}$ thresholds come from \citet{Pecaut2013} and are given in Table \ref{tab:Sp_type}. The spectral type is needed to more accurately determine the occurrence rate of planets, as discussed in Sect. \ref{sec:Planet population}.

\begin{table}[ht]
	\centering
	\caption{Spectral type classification as a function of $T_{eff}$.}
	\label{tab:Sp_type}
	\begin{tabular}{cc}
		\hline
		Spectral Type & $T_{eff}$ Range (K)\\
		\hline
		A & 7330 - 10050\\
        F & 5980 - 7330\\
        G & 5310 - 5980\\
        K & 3905 - 5310\\
        M & 2285 - 3905\\
		\hline
	\end{tabular}
\end{table}

\begin{figure}[ht]
	\includegraphics[width=\columnwidth]{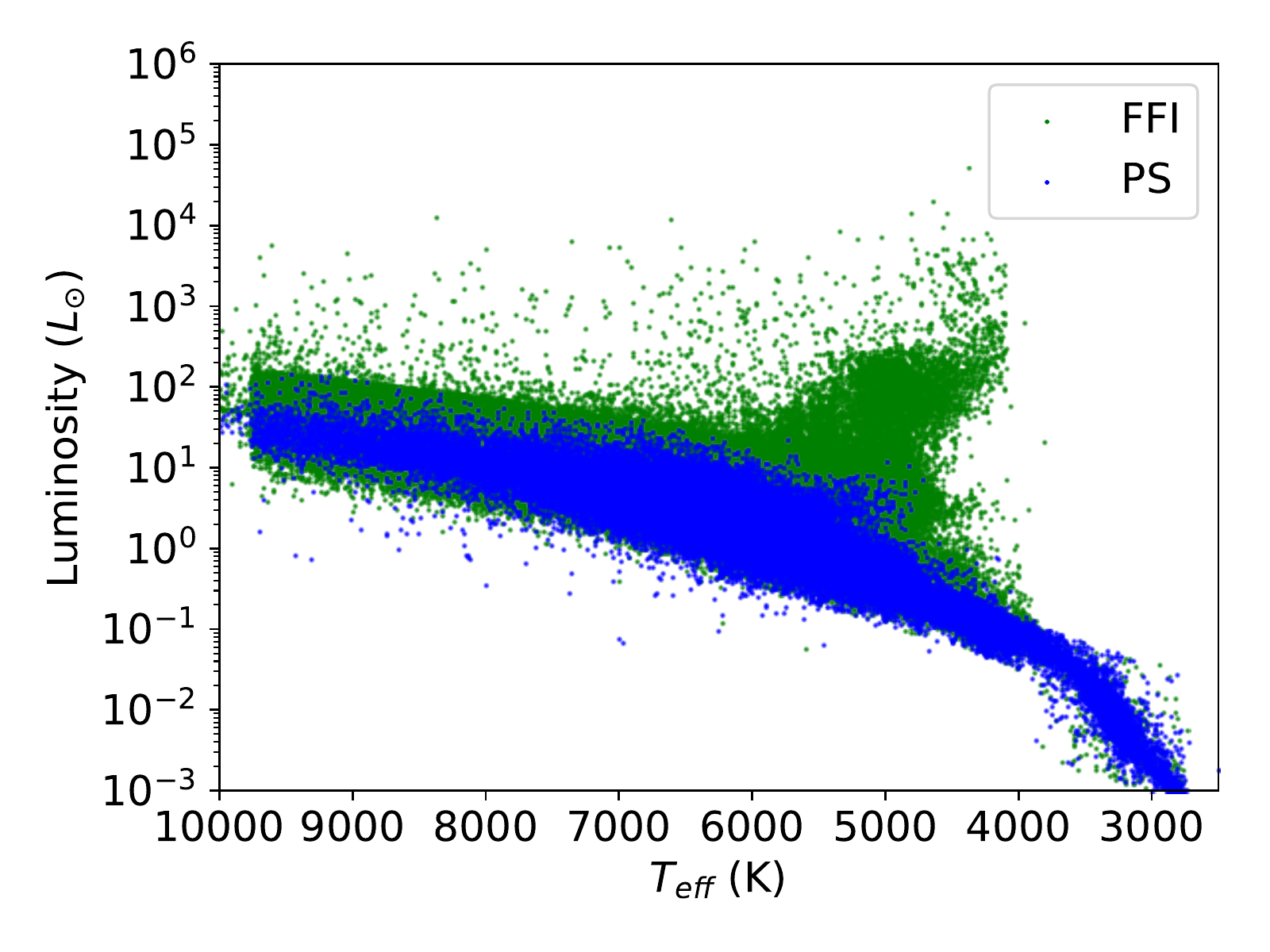}
    \caption{TIC CTL stellar distribution used in this simulation. Stars observed at 2-minute cadence (postage stamps) are shown in blue and stars observed at 30-minute cadence (full frame images) are in green.}
    \label{fig:HR}
\end{figure}

Figure \ref{fig:HR} shows the stellar distribution of \textit{TESS} targets on an HR diagram. The distribution is split up based on observing cadence used for each target (as described in Sect. \ref{sec:Detectability}).

\section{Planet population}
\label{sec:Planet population}

The simulated planet population was created by drawing from two distributions according to spectral type. The initial step was to determine the planet occurrence rates. The occurrence rate of planets, $\mu$, was taken to be 2.5 for M stars \citep{Dressing2015} and 0.689 for AFGK stars \citep{Fressin2013}. Taking \citet{Barclay2018} as a guide, the number of planets around each star was then predicted by drawing from a Poisson distribution centred on $\mu$ and is thus dependent on spectral type and therefore $T_{eff}$.

\begin{figure}[ht]
\centering
	\subfigure[Input distribution of planets with period.]{\label{fig:x}\includegraphics[width=\columnwidth]{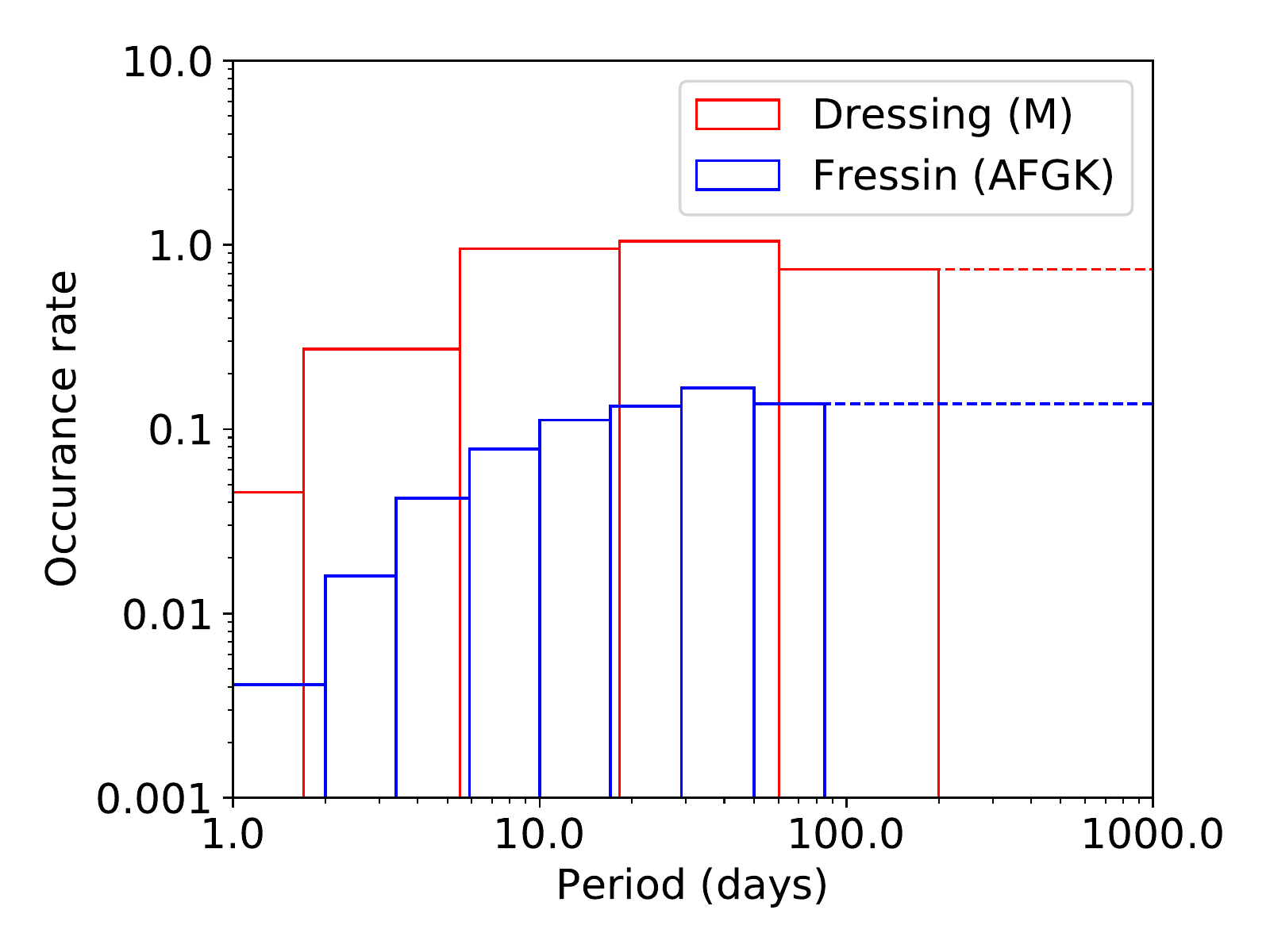}}
	\subfigure[Input distribution of planets with radius.]{\label{fig:y}\includegraphics[width=\columnwidth]{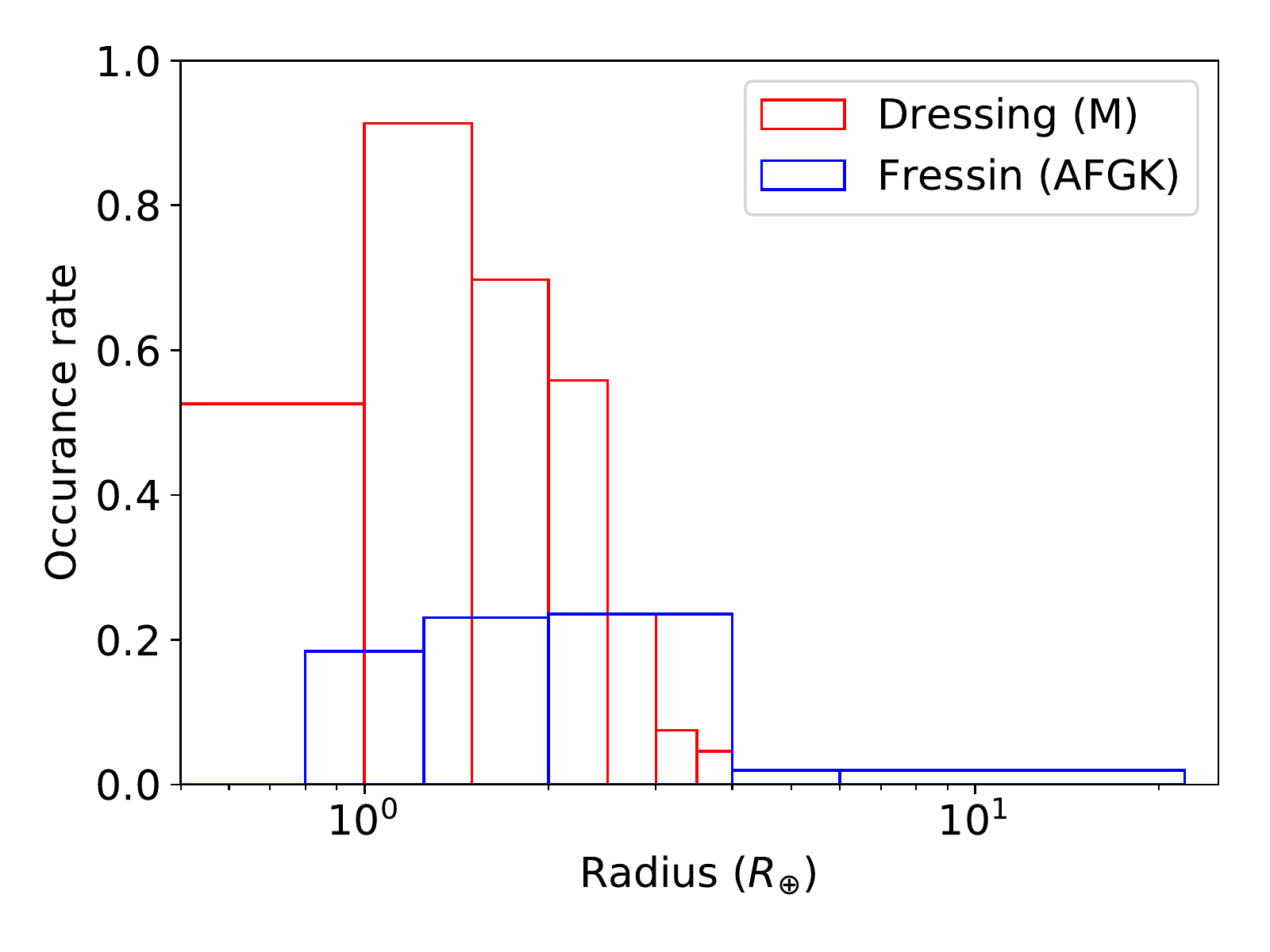}}
\caption{Input planet distributions used to simulated \textit{TESS} planets. The distribution for M stars is shown in red and AFGK stars are in blue. The dotted lines in Fig. \ref{fig:x} show the extension of the last period bins up to 1000 days.}
\label{fig:input_dist}
\end{figure}

Once the number of planets per star was decided the radius and period of these planets were drawn from two separate distributions depending on whether the star was an M type (in which case the distribution presented by \citet{Dressing2015} was employed) or an AFGK type (in which case the distribution presented by \citet{Fressin2013} was used). The distributions presented in these papers give results in terms of period/radius bins. To choose parameters a bin was chosen at random, weighted by the number of systems within that bin divided by the total number of systems in the sample. Once a bin was chosen the precise value of radius and period were chosen from a distribution between the limits of the bin. For radius the distribution was uniform between the bin limits whereas for period a uniform logarithmic distribution was used to give greater weighting to shorter period planets for which the period distributions are more complete \citep{Villanueva2018}. Figure \ref{fig:input_dist} shows the input distributions in period and radius from \citet{Dressing2015} and \citet{Fressin2013}. It should be noted that the input distributions used are only complete up to 85 and 200 days for AFGK and M stars respectively. To explore longer periods the final period bins were extended up to 1000 days (as demonstrated in \citet{Villanueva2018}).

For each simulated planet the orbital separation was then calculated using Kepler's third law,

\begin{equation}
\label{eq:orbital_sep}
a = \sqrt[3]{\frac{GM_{\star}P^2}{4\pi^2}}.
\end{equation}

\noindent
Next, values for the inclination, $i$, eccentricity, $e$, and periastron angle, $\omega$, were chosen. The periastron angle was drawn from a uniform distribution between $-\pi$ and $\pi$ whereas eccentricity was drawn from a beta distribution with $\alpha=1.03$ and $\beta=13.6$ which has been found to be a valid assumption for transiting exoplanet populations \citep{VanEylen2015}. Inclination was stored as $\cos{i}$ and is chosen from a uniform distribution between $0$ and $1$. Inclination is assumed to be the same for all planets in the same system. Finally, the transit duration $T_{dur}$ is calculated using equation (1) from \citet{Barclay2018};

\begin{equation}
\label{eq:t_dur}
T_{dur} = \frac{P}{\pi} \times \arcsin\left({\frac{R_{\star}}{a} \times \frac{\sqrt{1 + \frac{r_p}{R_{\star}} - b^2}}{\sqrt{1 - \cos^2{i}}}}\right).
\end{equation}

\section{Detectability}
\label{sec:Detectability}

To determine whether a planet will have a detectable transit during its observations by \textit{TESS} it must first be determined whether the planet transits at all. To determine this the impact parameter is calculated using an equation from \citet{Winn2014} reproduced here,

\begin{equation}
\label{eq:impact_param}
b = \frac{a\cos{i}}{R_{\star}} \times \frac{1-e^2}{1+e\sin{\omega}}.
\end{equation}

\noindent
A transit is defined as occurring for $\lvert b \rvert<1$. This is a good approximation for the majority of systems excluding large planets around M stars. However, since these systems are relatively rare \citep{Dressing2015} the approximation can be used without significant effects on the simulation outcome \citep{Barclay2018}.

Next it must be determined whether a transit will occur during the observation of its host. A random number, designated $T_0$, is generated from $0$ to $P$ (where $P$ is the orbital period). If $T_0 < T_{obs}$ a transit will be observed. Here $T_{obs}$ is the length of time for which a planet system is observed by \textit{TESS}. Further comparison of $T_0$ to $T_{obs}$, accounting for period, $P$, shows how many transits are observed during the mission, and this is denoted as $n$. The final step is then to determine whether the transit signal will be detectable above the noise. The S/N is given by

\begin{equation}
\label{eq:SNR}
S/N = \frac{\delta_{eff}}{\sigma_{1hr}}\sqrt{\frac{T_{dur}}{\Delta T}}\sqrt{n}
\end{equation}

\noindent
where $\delta_{eff}$ is effective transit depth, $\sigma_{1hr}$ is the total noise measured in 1 hour of data, $n$ is number of observed transits, $T_{dur}$ is transit duration in hours and $\Delta T$ is observing cadence. The factor of $\Delta T$ is included to account for different levels of noise that will be present depending on the cadence of the observations. To be consistent with previous work this is the same S/N prescription as employed by \citet{Villanueva2018} and \citet{Barclay2018}.

\begin{figure}[ht]
	\includegraphics[width=\columnwidth]{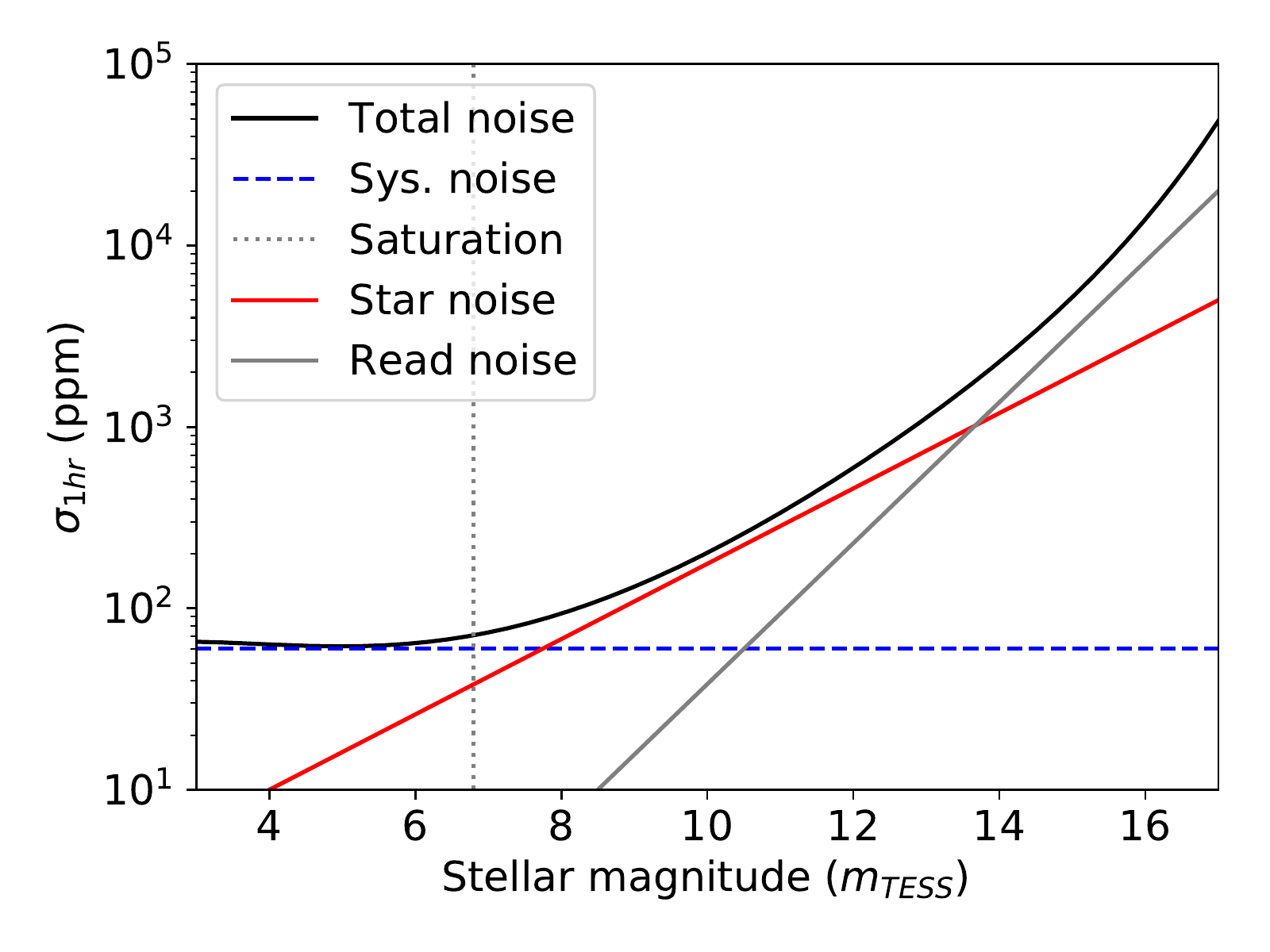}
    \caption{Total noise in 1 hour as a function of \textit{TESS} magnitude calculated using equation \ref{eq:noise}. Also shown are individual contributions from star noise, read noise and systematic noise as well as a predicted saturation cut off \citep{Sullivan2015}. It is noted that the total noise budget will saturate at 60ppm.}
    \label{fig:noise}
\end{figure}

Following the example of \citet{Stassun2017} we calculate the noise in 1hr using a fifth-order polynomial approximation,

\begin{equation}
\begin{split}
\label{eq:noise}
\ln{\left(\sigma_{1hr}\right)} = C_1 + C_2m_{TESS} + C_3m_{TESS}^2 + C_4m_{TESS}^3 +\\ C_5m_{TESS}^4 + C_6m_{TESS}^5,
\end{split}
\end{equation}

\noindent
which is a direct function of the \textit{TESS} magnitude, $m_{TESS}$. The constants are given as $C_1 = 3.9269$, $C_2 = 0.8500$, $C_3 = -0.2850$, $C_4 = 0.0396$, $C_5 = -0.0022$ and $C_6 = 4.7351\times10^{-5}$. Figure \ref{fig:noise} shows the noise profile employed here (for $3 \leq m_{TESS} \leq 17$) with some of the contributing sources.

The effective transit depth $\delta_{eff}$ is found using

\begin{equation}
\label{eq:delta}
\delta_{eff} = \left(\frac{R_p}{R_{\star}}\right)^2 \times \frac{1}{1+C},
\end{equation}

where $C$ denotes the contamination ratio from background stars, defined as the ratio of background flux to source flux, $C=F_b/F_s$ \citep{Stassun2017}. Finally, it must be decided whether each target is observed a 2-minute cadence (Postage Stamps) or 30-minute cadence (Full Frame Images) \citep{Ricker2015}. The choice of cadence value is based on a priority metric;

\begin{equation}
\label{eq:metric}
\frac{\sqrt{N_s}}{\sigma_{1hr} R_{\star}^{3/2}}
\end{equation}

where $N_s$ is the number of sectors for which a target is observed, $\sigma_{1hr}$ is the noise in 1hr and $R_{\star}$ is the stellar radius. This priority metric is provided in the CTL but we recalculated it in this paper to allow for changes resulting from the observing sector locations. The top ranked 200,000 stars are then assigned 2-minute cadence with all others being observed at a cadence of 30-minutes. From equation \ref{eq:metric} is is clear that 2-minute cadence targets are biased towards the \textit{TESS} CVZ through the factor of $\sqrt{N_s}$.

A transit is determined to be detectable when it has $S/N>7.3$, a value chosen so as to be consistent with the current literature \citep[e.g.][]{Sullivan2015,Bouma2017,Barclay2018,Villanueva2018}.

Table \ref{tab:numbers table} shows some of the actual numbers of various subsets of the simulated planets. These numbers are determined as the mean of multiple simulation runs with the uncertainties defined as the standard deviation of all runs. Key in these numbers is the separation based on the number of \textit{TESS} sectors which will observe each potential host. From this table (as well as from Fig. \ref{fig:sector analysis}) it is seen that the number of stars within each region is almost exactly proportional to the area of sky covered by that region. As before, the minor differences are due to the CTL stars not being exactly uniform in their distribution across the celestial sphere. The values presented in Table \ref{tab:numbers table} show that approximately 4 million planets are simulated with 4721$\pm$181 being detectable by \textit{TESS} and 464$\pm$41 being both detectable and single transiting. These numbers are further broken down to show the contributions from the 2-minute and 30-minute cadence targets. This is a vital statistic as the cadence may heavily influence the accuracy with which a single transit signal can be modelled, which in turn, influences the reliability of follow-up efforts. For a discussion of the prediction of orbital parameters based on single transit events see \citet{Osborn2015}. Note that due to the factor of $\sqrt{N_s}$ in equation \ref{eq:metric} there are no 2-minute cadence stars in the area of the sky not observed by \textit{TESS}. Obviously, the numbers presented here are dependent on \textit{TESS} working to specifications. Any departure from this, for example increased background due to scattered light, would affect the detection rate of planets, especially those around fainter hosts.

\begin{table*}[ht]
\centering
\caption{Numbers of simulated planets. This table displays planets by number of sectors that observe them as well as making splits based on detectability, single transit nature and observing cadence. Targets observed by 5-11 sectors are combined as individually they comprise very small fractions of the total. Also included is the percentage of sky area observed by each number of sectors and the percentage of total simulated planets in each region. The given values are averaged from multiple simulation runs.}
\label{tab:numbers table}
\begin{tabular}{|c||c|c||c|c||c|c||c|c|}
\hline
{Number Of Sectors} & \multicolumn{2}{c||}{All Planets} & \multicolumn{2}{c||}{Detectable Planets} & \multicolumn{2}{c||}{Detectable Single Transits} & {Sky Area} & {Planets}\\ \cline{2-7}
Observed By (Baseline)                       & 2 min          & 30 min          & 2 min              & 30 min             & 2 min                & 30 min                &            (\%) & (\%)     \\ \hline\hline
0 (0.0)                & 0$\pm$0        & 583049$\pm$374   & 0$\pm$0     & 0$\pm$0     & 0$\pm$0   & 0$\pm$0    & 14.60  & 14.29  \\ \hline
1 (27.4)               & 101633$\pm$158 & 2546775$\pm$1445 & 535$\pm$27  & 2231$\pm$38 & 92$\pm$9  & 317$\pm$18 & 63.00  & 65.01  \\ \hline
2 (54.8)               & 45123$\pm$226  & 551382$\pm$352   & 289$\pm$15  & 737$\pm$21  & 15$\pm$4  & 35$\pm$6   & 15.20  & 14.61  \\ \hline
3 (82.2)               & 13083$\pm$73   & 89545$\pm$265    & 92$\pm$5    & 158$\pm$10  & 1$\pm$1   & 3$\pm$1    & 3.00   & 2.52   \\ \hline
4 (109.6)              & 3790$\pm$57    & 18629$\pm$87     & 30$\pm$6    & 35$\pm$7    & 0$\pm$0   & 1$\pm$1    & 0.56   & 0.55   \\ \hline
5 – 11 (137.0 – 301.4) & 14256$\pm$87   & 35008$\pm$85     & 119$\pm$8   & 73$\pm$9    & 0$\pm$1   & 0$\pm$0    & 1.40   & 1.20   \\ \hline
12 (328.8)             & 397$\pm$7      & 540$\pm$18       & 4$\pm$1     & 1$\pm$1     & 0$\pm$0   & 0$\pm$0    & 0.52   & 0.02   \\ \hline
13 (356.2)             & 34925$\pm$208  & 38586$\pm$173    & 338$\pm$27  & 80$\pm$6    & 0$\pm$0   & 0$\pm$0    & 1.70   & 1.80   \\ \hline\hline
Cadence Total          & 213236$\pm$815 & 3863513$\pm$2799 & 1407$\pm$89 & 3314$\pm$92 & 108$\pm$15 & 356$\pm$26 & 100.00 & 100.00  \\ \hline
Full Total                                & \multicolumn{2}{c||}{4076748$\pm$3614}     & \multicolumn{2}{c||}{4721$\pm$181}              & \multicolumn{2}{c||}{464$\pm$41}                    &    \multicolumn{2}{c|}{}                       \\ \hline
\end{tabular}
\end{table*}

\begin{figure}[ht]
	\includegraphics[width=\columnwidth]{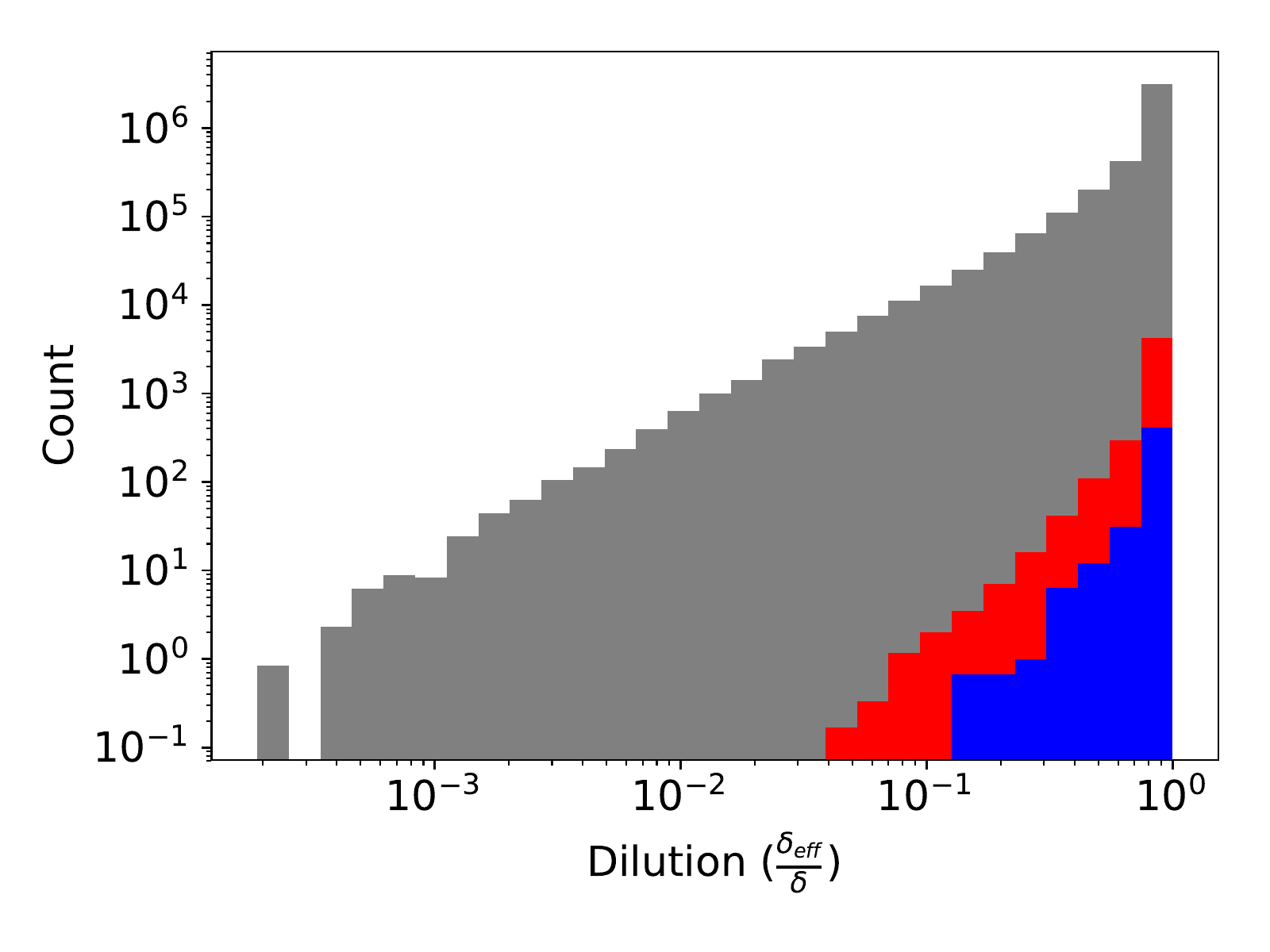}
    \caption{Plot of dilution of simulated planet hosts. Lower values correspond to more diluted signals.}
    \label{fig:contratio}
\end{figure}

It is also key to note the effect of contamination ratio, $C$, on the detectability of targets. Since \textit{TESS} pixels are relatively large ($21\times21$ arcsec/pixel) contamination is a non trivial issue \citep{Stassun2017}. Figure \ref{fig:contratio} shows a plot of transit depth, $\delta$, against effective transit depth, $\delta_{eff}$, showing the impact of background contamination.

From this Fig. \ref{fig:contratio} it is seen that the majority of detections are found around weakly contaminated hosts meaning that $\delta \approx \delta_{eff}$. Therefore, it is obvious that efforts to reduce, or correct for, contamination effects may lead to large increases in the number of exoplanet detections by \textit{TESS}.

\section{Single site follow-up}
\label{sec:Single site follow-up}

\begin{figure}[ht]
	\includegraphics[width=\columnwidth]{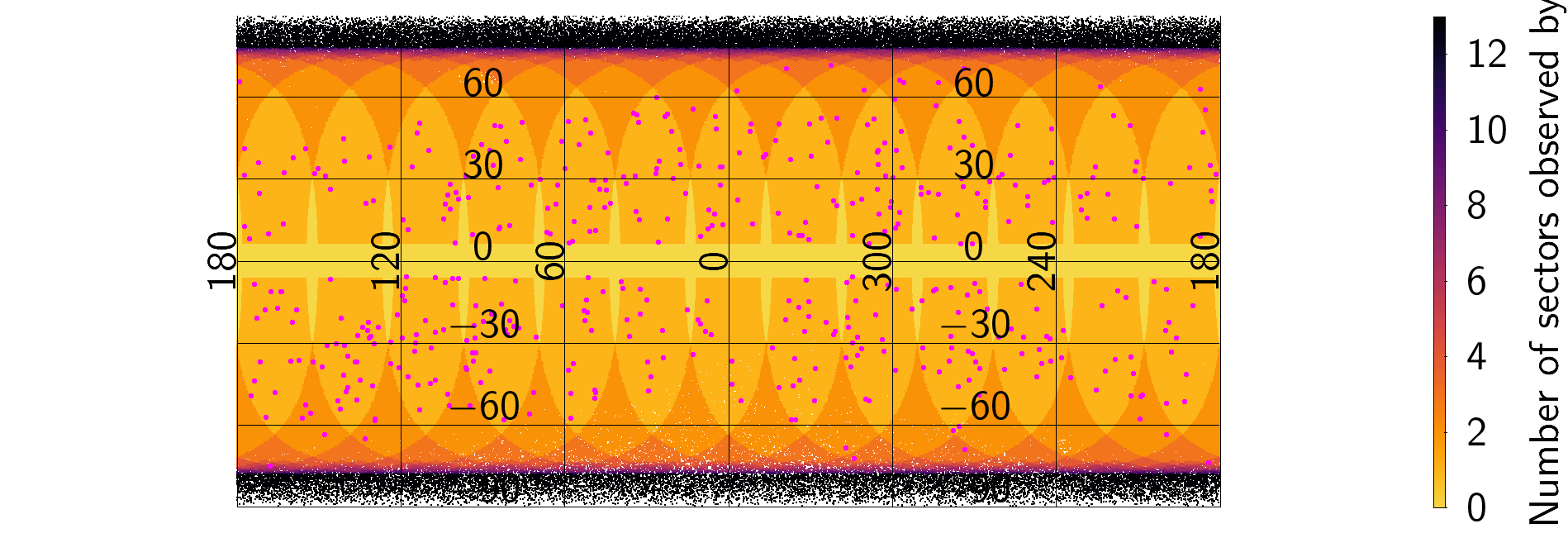}
    \caption{Plot of all simulated planets coloured by number of sectors observed for. The detectable single transits for a single simulation run (490 targets) are overlaid in pink.}
    \label{fig:singletrans}
\end{figure}

After simulating all detectable transiting planets around CTL stars cuts were made requiring $n=1$ to select only single transit detections. Figure \ref{fig:singletrans} shows a plot of all simulated planets with an overlay of these single transits on an ecliptic sky map.

Next, the dates at which \textit{TESS} will observe these transits are found. The dates are calculated relative to the beginning of \textit{TESS} observations and are then arbitrarily offset to begin at the start of 2019. Once the dates and times of \textit{TESS} single transits are found it must be determined which events can be observed from a single ground-based site. Here the chosen site is Paranal, Chile but the process can be generally extended to any desired site with known geographical coordinates. This is achieved by predicting the times of central transit using our custom tool. Obviously, this process assumes that the epoch and period of a transit event can be accurately predicted based on a single \textit{TESS} light curve. In reality there will be uncertainties in these measurements meaning that a target will most likely need to be observed for some time prior and post predicted transit time to ensure the transit is observed while allowing for uncertainties in ephemeris, this is further discussed in Sec. \ref{sec:Realistic follow-up}.

An analysis of the improvement of these ephemeris uncertainties is carried out by \citet{Yao2018}. This paper simulates \textit{TESS} single transit signals and injects them into \textit{KELT} light curves. The paper then determines how many of the signals are recoverable from the \textit{KELT} light curves and finds that the number of recoverable signals is dependent on the planet radius and period with up to 80\% being recoverable for the largest radius planets, falling to 18\% for the smallest. For signals not recoverable in the \textit{KELT} data the period uncertainty will remain around 10\% \citep{Osborn2015} but for signals that can be seen in \textit{KELT} the additional data will better constrain the ephemeris.

\section{Analysis and Results}
\label{sec:Analysis and Results}

\subsection{\textit{TESS} single transit demographics}
\label{sec:TESS single transit demographics}

\begin{figure*}[ht]
\centering
	\subfigure[]{\label{fig:a}		\includegraphics[width=0.45\linewidth]{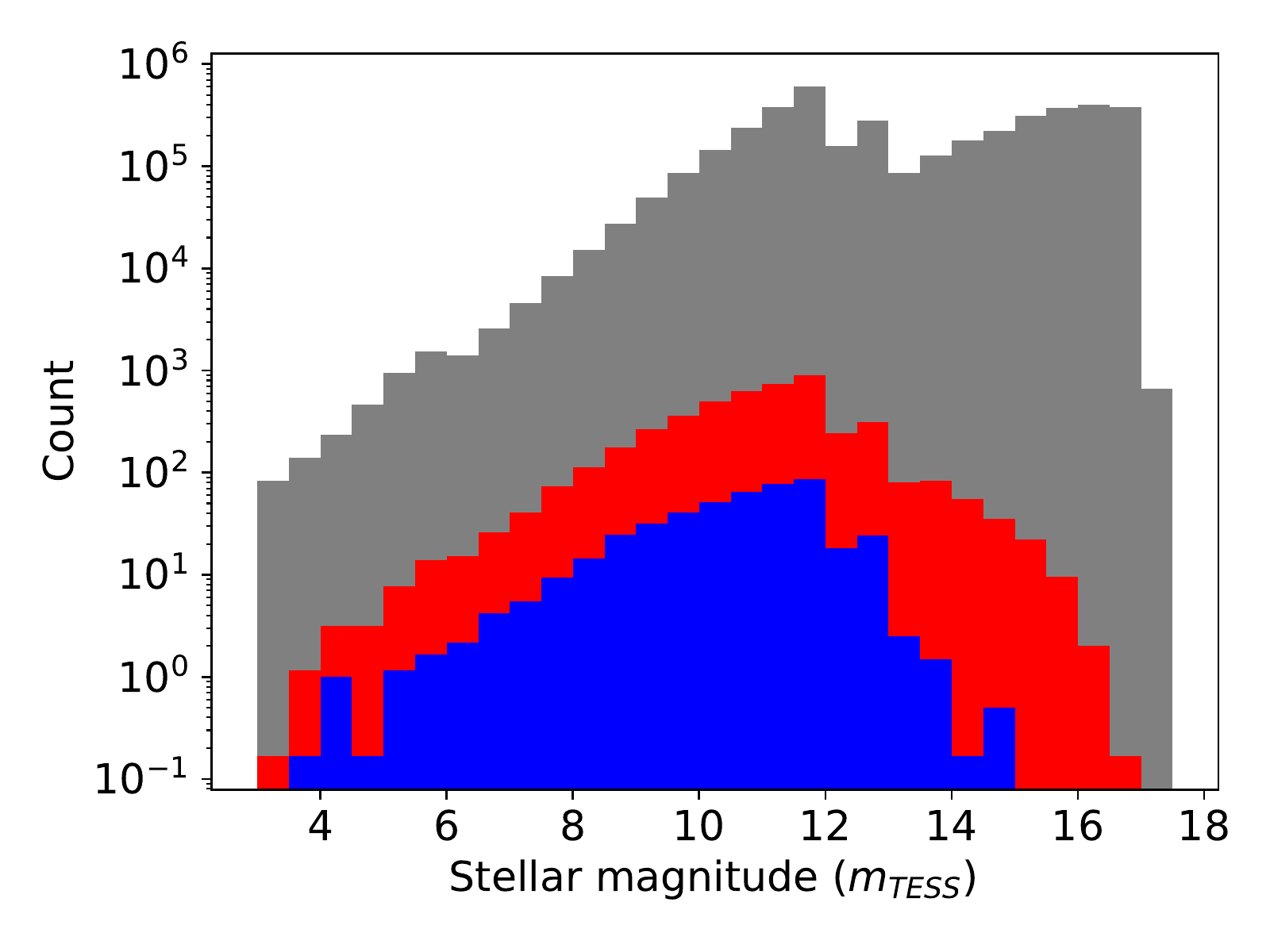}}
	\subfigure[]{\label{fig:b}\includegraphics[width=0.45\linewidth]{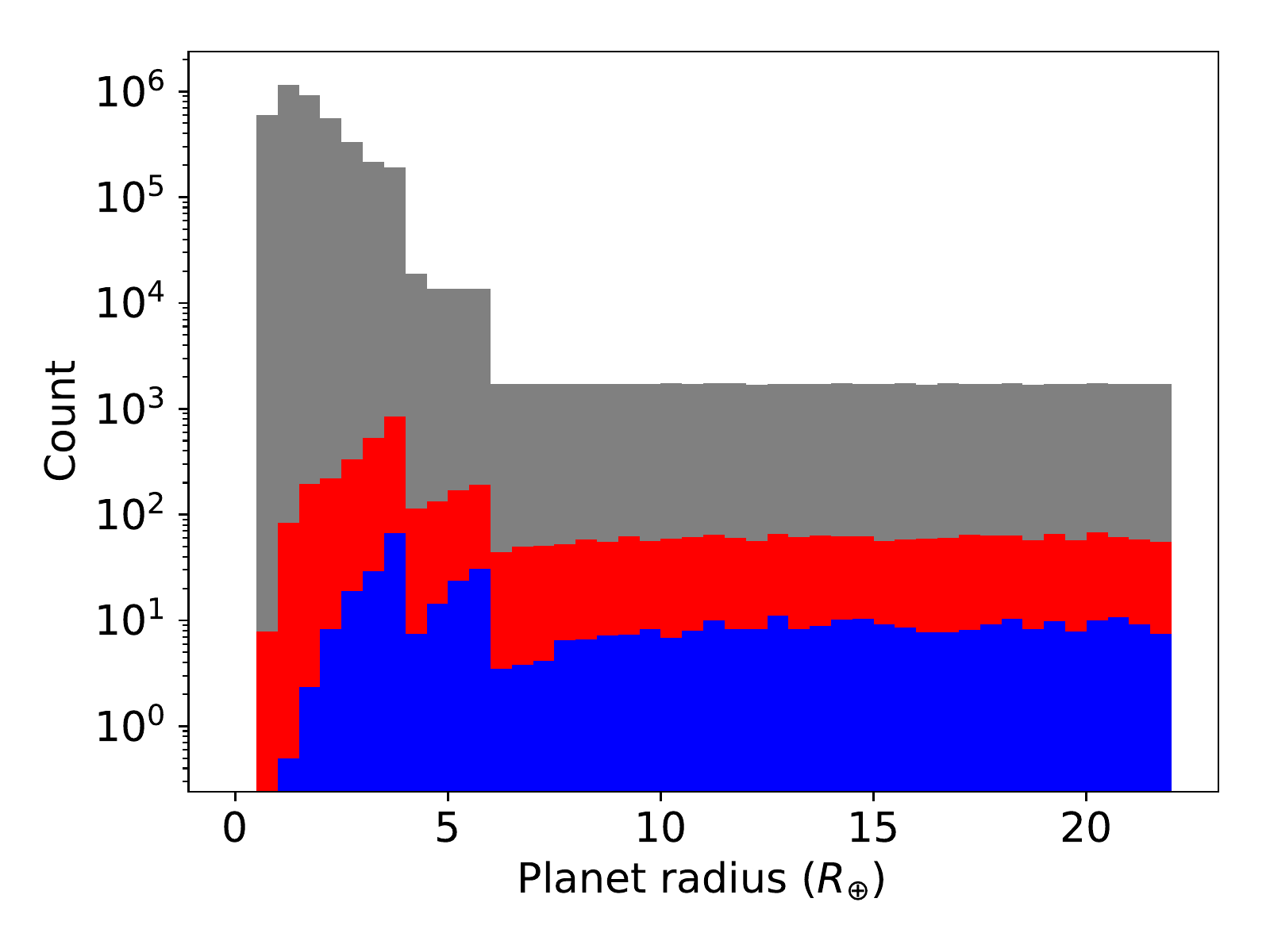}}
	\subfigure[]{\label{fig:c}		\includegraphics[width=0.45\linewidth]{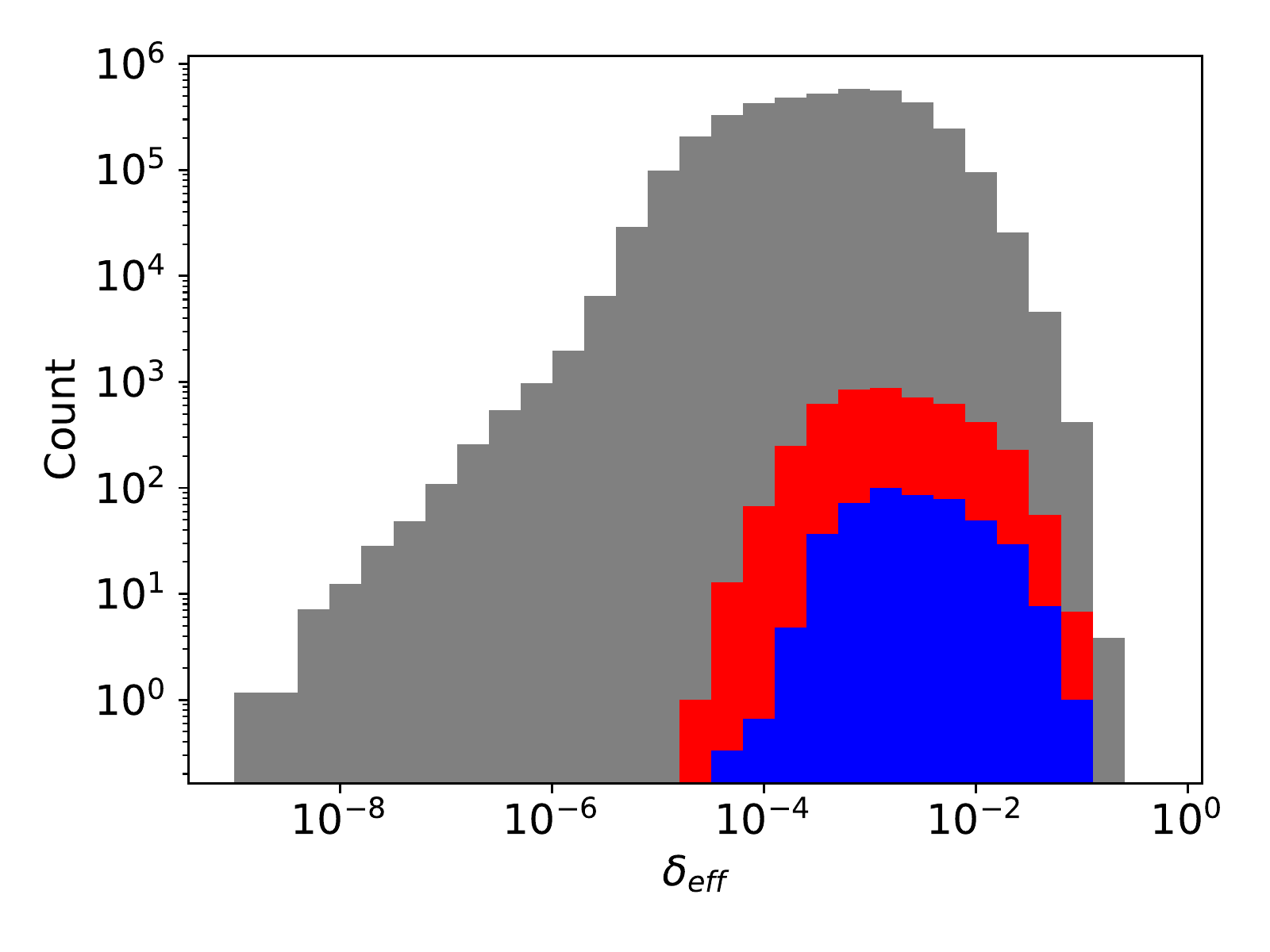}}
	\subfigure[]{\label{fig:d}\includegraphics[width=0.45\linewidth]{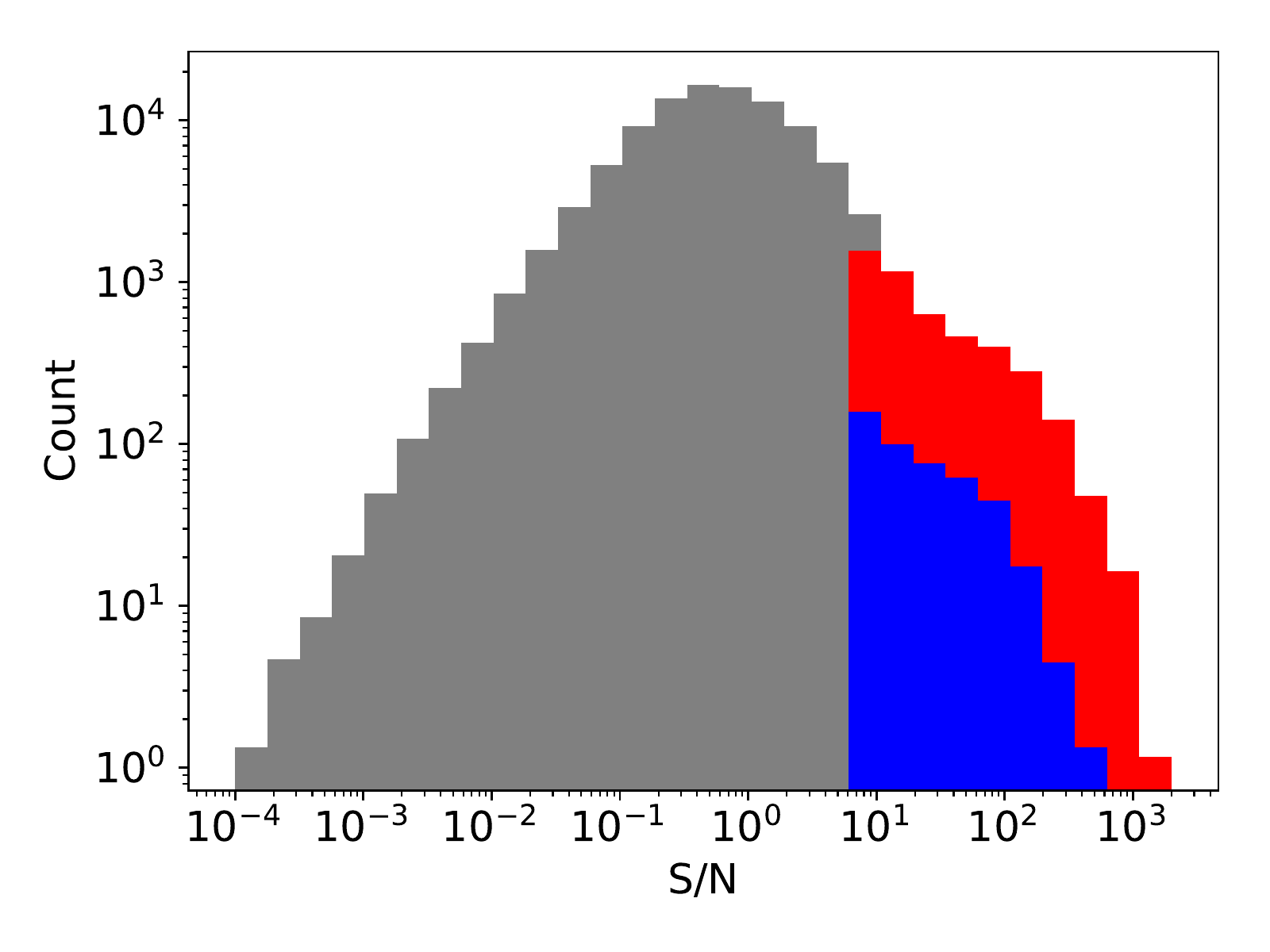}}
\caption{Simulated planet demographics. Grey shows all simulated planets, red shows all planets for which a transit is observed at $S/N>7.3$ and blue shows all single transits with $S/N>7.3$.}
\label{fig:planet_dist}
\end{figure*}

\begin{figure}[ht]
\centering
	\subfigure[Orbital period distribution for all simulated planets]{\label{fig:i}		\includegraphics[width=\columnwidth]{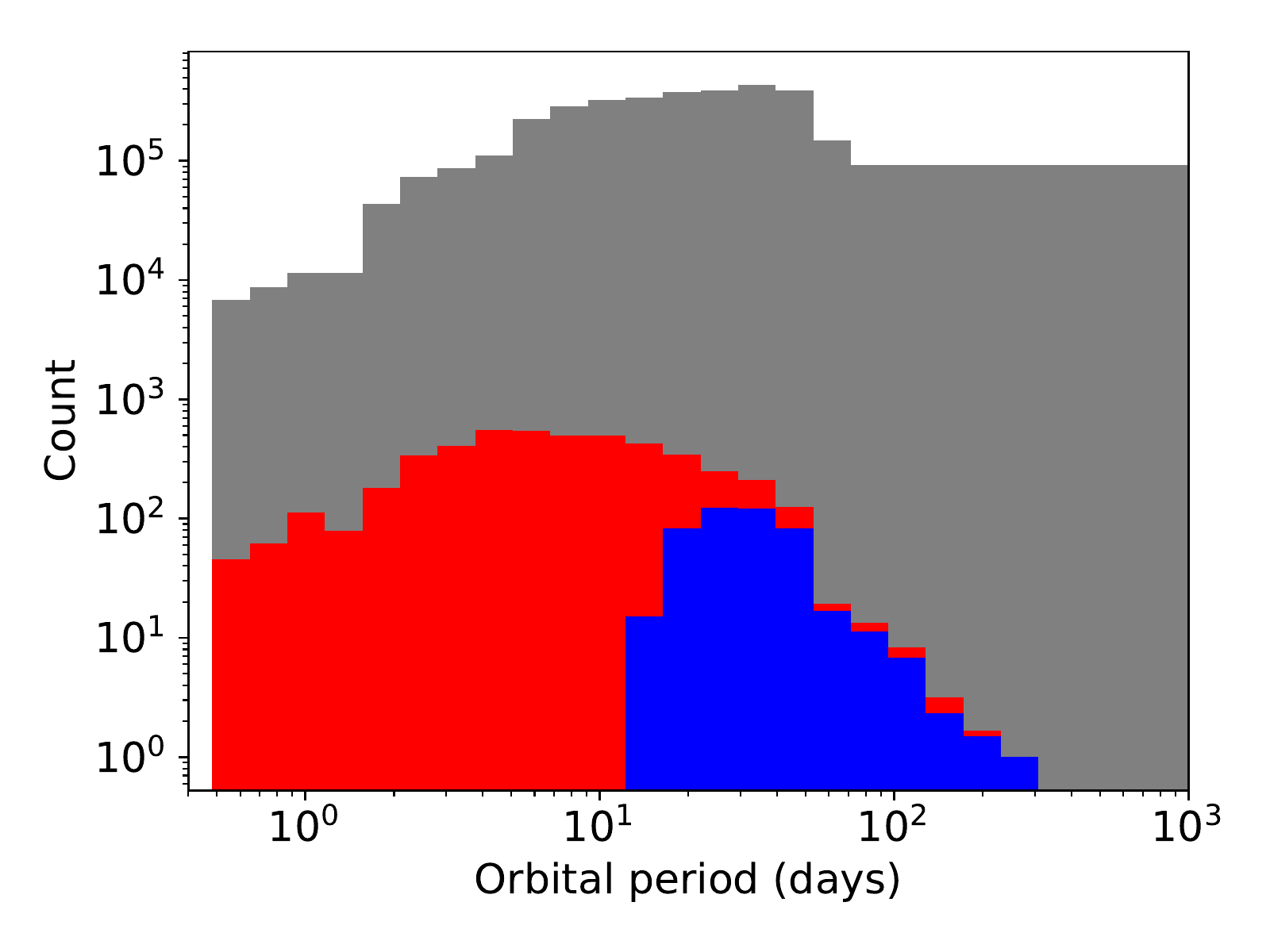}}
	\subfigure[Orbital period distribution for all detectable planets with $P>25$ days.]{\label{fig:ii}\includegraphics[width=\columnwidth]{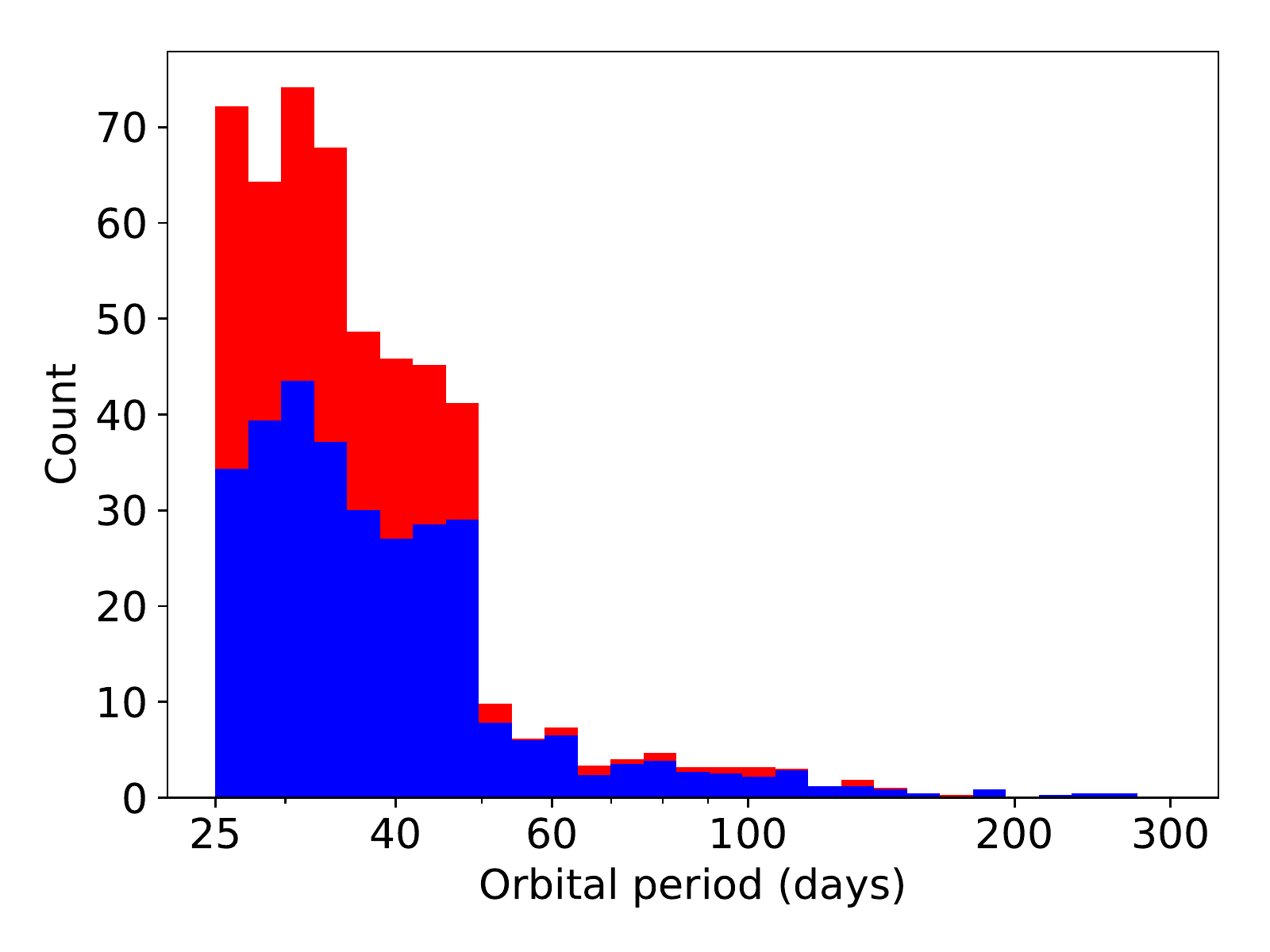}}
\caption{Simulated planet orbital periods averaged over multiple simulation runs. Colours are as in Fig. \ref{fig:planet_dist}. Undetectable planets are omitted from figure \ref{fig:ii} as they overwhelm the long period detectable planets.}
\label{fig:period}
\end{figure}

In Figs. \ref{fig:planet_dist} and \ref{fig:period} we show some of the demographics represented in the simulated planets around CTL stars. In each plot the distribution of all simulated planets is shown as well as the sub-distributions corresponding to all planets for which a transit is observed at $S/N>7.3$ and all single transits which are observed at $S/N>7.3$. The four plots shown in Fig. \ref{fig:planet_dist} show the planet distributions as functions of \textit{TESS} magnitude, planet radius, effective transit depth and S/N. The distributions shown in each subplot are the result of averaging multiple simulation runs.

The distribution with \textit{TESS} magnitude is seen in Fig. \ref{fig:a}. For bright stars ($m_{TESS} \leq 12$) the number of both multiple and single transit detections increases with magnitude. This is a function of the increased numbers of stars with larger magnitude values (approximately an exponential increase). For these bright stars the number of detections and single transit detections comprise approximately the same fraction of total planets at each magnitude ($\sim1\%$ and $\sim0.1\%$ respectively). For dimmer stars however, this fraction falls off rapidly. This is partially due to a decrease in CTL stars with $12<m_{TESS}<14$, but mainly due to the increased difficulty with which transits can be detected around dimmer stars, a direct consequence of the effect of magnitude on noise shown in equation \ref{eq:noise} which leads to a decrease in $S/N$.

Figure \ref{fig:b} shows the distribution as a function of planet radius. The overall distribution of simulated planets (in grey) is dictated by the input distribution seen in Fig. \ref{fig:x}. At large planet radii the number of detectable transits is proportional to total number of planets with approximately 3.5\% of planets being detectable and 0.5\% being detectable single transits. Below $\sim6R_{\oplus}$ however, this is no longer true. Though the number of simulated planets increases as the planet radius falls the number of detections remains constant, thus the detection rate falls. This is a simple consequence of the fact that smaller radius planets produce smaller transit signals which are harder to detect above to noise.

Figure \ref{fig:c} shows an important distribution. The pattern here shows that detectable transits peak with an effective transit depth at $\sim0.001$, this pattern is closely matched by the single transits. The effective transit depth here is only relevant in its relation to the photometric precision of \textit{TESS}. For follow-up however, transit depths that are too small will not be observable from ground based instruments which will have higher noise thresholds than space-based missions like \textit{TESS}. When looking specifically at the detection of single transits, \cite{Osborn2015} suggest that signals at the level of 1 mmag are potentially discoverable. Thus, we predict that the number of detectable single transits found by \textit{TESS} with effective depth $\delta_{eff}\geq0.001$ will be 350$\pm$25.

Finally, Fig. \ref{fig:d} shows the distribution of simulated planets with $S/N$. The limit of $S/N>7.3$ for a detection used here may seem overly optimistic. From this figure it can be seen that non-negligible numbers of single transits will be detected up to much larger $S/N$ values. For example, the number of detectable single transits with $S/N\geq20$ is 202$\pm$15.

An additional key demographic is the distribution of orbital period shown in Fig. \ref{fig:period}. Figure \ref{fig:i} shows the distribution of planets with orbital period. The distribution of all simulated planets (in grey) is due to the input distributions employed from \citet{Dressing2015} and \citet{Fressin2013} (seen in Fig. \ref{fig:y}). The number of detectable transits shows a similar distribution. There are obviously no single transits below $\sim15$ days as these would all be observed for a minimum of 2 transits, even if only seen in a single \textit{TESS} observing sector (with a baseline of 27.4 days). Figure \ref{fig:ii} shows a close up of periods with $P\geq25$ days. From this figure it can be seen that between 25 and 50 days approximately 70\% of all detections will come from single transits with the number of single transit detections falling off slower than the total number of detections. Above 50 days this fraction approaches 100\%. The  number of single transits detectable by \textit{TESS} with $P\geq25$ days is found to be 315$\pm$22. For periods longer than 50 days the number is 44$\pm$7. The full break down of detections with period is shown in Table \ref{tab:period table}. This table shows the number of \textit{TESS} detections in 20 period bins, looking at multiple and single transit detections separately. Also shown in Table \ref{tab:period table} is the sub-set of detections which will be found with sub-Neptune radii ($R_p \leq 4R_{\oplus}$).

\begin{table}[ht]
\centering
\caption{Number of detectable planets found in each of 20 logarithmic period bins. Planets are separated into multitransiting planets and single transiting planets. Also shown is the number of sub-Neptune planets ($R_p \leq 4R_{\oplus}$). Note that planets with $P<1$ day are omitted. Given values and uncertainties are the result of averaging multiple simulation runs.}
\label{tab:period table}
\resizebox{\columnwidth}{!}{\begin{tabular}{|c|c|c|}
\hline
Period Bin (days) & Multitransits ($R_p \leq 4R_{\oplus}$) & Single Transits ($R_p \leq 4R_{\oplus}$) \\ \hline\hline
1.0 - 1.4         & 107$\pm$8  (59$\pm$5)        & 0$\pm$0    (0$\pm$0)         \\ \hline
1.4 - 2.0         & 135$\pm$18 (87$\pm$13)       & 0$\pm$0    (0$\pm$0)         \\ \hline
2.0 - 2.8         & 414$\pm$20 (212$\pm$15)      & 0$\pm$0    (0$\pm$0)         \\ \hline
2.8 - 4.0         & 518$\pm$29 (216$\pm$16)      & 0$\pm$0    (0$\pm$0)         \\ \hline
4.0 - 5.6         & 601$\pm$26 (219$\pm$10)      & 0$\pm$0    (0$\pm$0)         \\ \hline
5.6 - 7.9         & 658$\pm$18 (354$\pm$15)      & 0$\pm$0    (0$\pm$0)         \\ \hline
7.9 - 11.2        & 574$\pm$15 (290$\pm$10)      & 0$\pm$0    (0$\pm$0)         \\ \hline
11.2 - 15.8       & 525$\pm$23 (234$\pm$23)      & 11$\pm$3   (3$\pm$1)         \\ \hline
15.8 - 22.4       & 307$\pm$20 (159$\pm$8)       & 95$\pm$12  (27$\pm$4)        \\ \hline
22.4 - 31.6       & 141$\pm$8  (77$\pm$10)       & 150$\pm$10 (47$\pm$5)        \\ \hline
31.6 - 44.7       & 88$\pm$5   (46$\pm$4)        & 127$\pm$12 (35$\pm$5)        \\ \hline
44.7 - 63.1       & 19$\pm$4   (10$\pm$2)        & 54$\pm$7   (12$\pm$3)        \\ \hline
63.1 - 89.1       & 3$\pm$1    (2$\pm$1)         & 13$\pm$4   (1$\pm$1)         \\ \hline
89.1 - 125.9      & 2$\pm$1    (1$\pm$1)         & 9$\pm$4    (1$\pm$1)         \\ \hline
125.9 - 177.9     & 1$\pm$1    (0$\pm$0)         & 3$\pm$1    (0$\pm$1)         \\ \hline
177.9 - 251.2     & 0$\pm$0    (0$\pm$0)         & 2$\pm$1    (0$\pm$0)         \\ \hline
251.2 - 354.8     & 0$\pm$0    (0$\pm$0)         & 1$\pm$1    (0$\pm$0)         \\ \hline
354.8 - 501.2     & 0$\pm$0    (0$\pm$0)         & 0$\pm$0    (0$\pm$0)         \\ \hline
501.2 - 707.9     & 0$\pm$0    (0$\pm$0)         & 0$\pm$0    (0$\pm$0)         \\ \hline
707.9 - 1000.0    & 0$\pm$0    (0$\pm$0)         & 0$\pm$0    (0$\pm$0)         \\ \hline\hline
Total             & 4093$\pm$200 (1965$\pm$133)  & 464$\pm$55 (126$\pm$22)      \\ \hline
\end{tabular}}
\end{table}

\begin{figure}[ht]
	\includegraphics[width=\columnwidth]{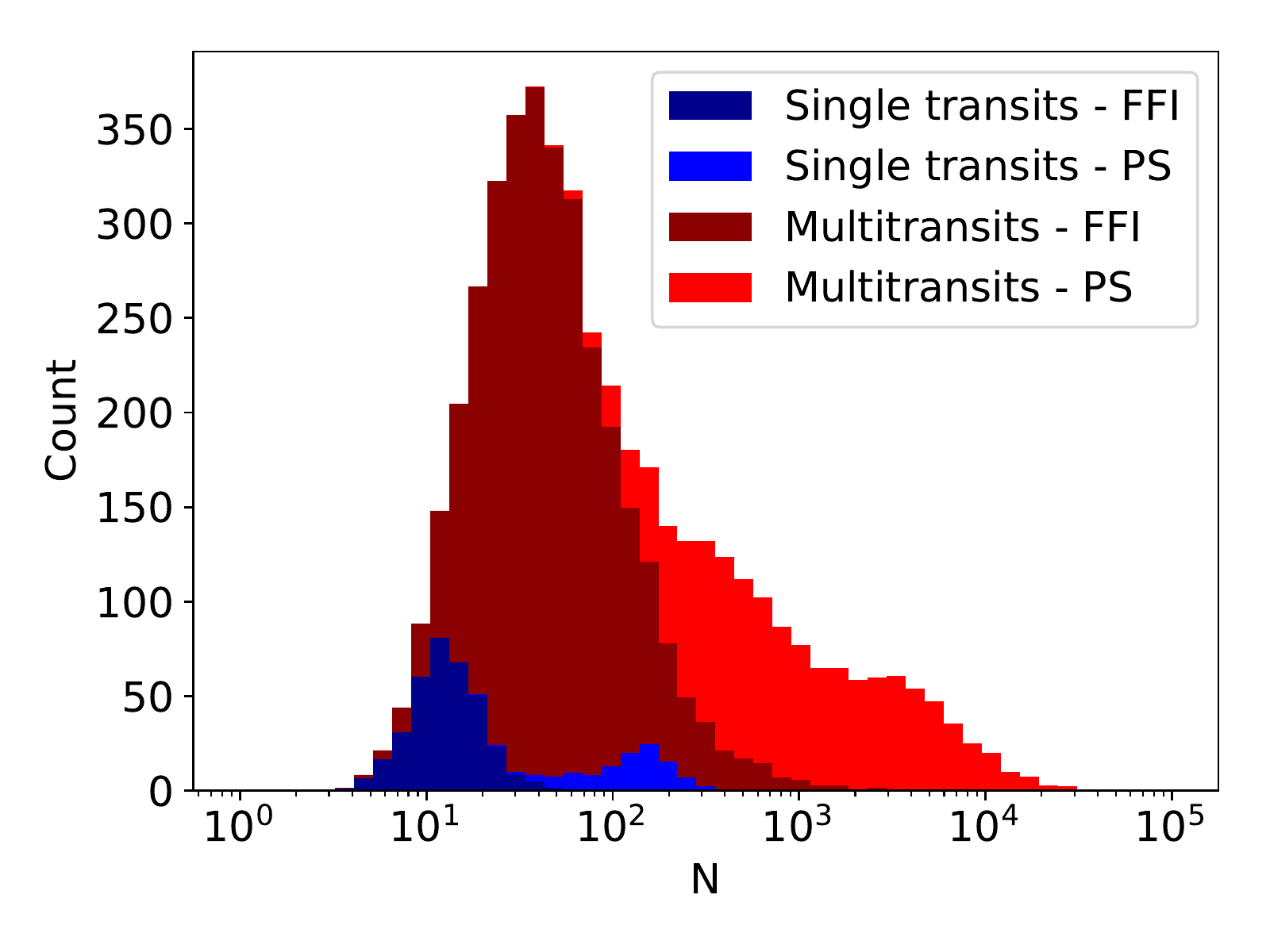}
    \caption{Distribution of detectable planets with $N$, averaged from multiple simulation runs. This plot shows the distribution broken down by number of transits observed (multiple or single transits) and observing cadence (2 or 30 minutes). The four distributions are stacked so the maximum height of each bar shows the total detections for each $N$.}
    \label{fig:N}
\end{figure}

Additionally important, especially for follow-up of candidates, is the accuracy with which planetary and orbital parameters can be predicted from a single observed transit \citep{Osborn2015}. To better constrain these values requires more in transit observations. The number of in transit observations, $N$, is given by

\begin{equation}
\label{eq:N}
N = \frac{T_{dur}}{\Delta T}\times n,
\end{equation}

where $\Delta T$ is the cadence at which a given host is observed and $n$ is the number of observed transits. Therefore, Fig. \ref{fig:N} shows how the detectable planets break down as a function of the number of in transit observations, $N$.

The red distributions in Fig. \ref{fig:N} show the multitransit detections with the 2-minute cadence targets in the darker shade of red. The blue distributions show the single transit detections with the 2-minute cadence targets in the darker shade of blue. It can be seen that there is a very obvious bi-modality in the distribution based on observing cadence. This is to be expected. Observations at 2-minute cadence will have, assuming identical values of transit duration, 15 times more in transit observations; $N_{2min} \approx 15N_{30min}$. This effect is seen in both multiple and single transits but is less obvious for multitransits as additional transits also contribute to increasing the value of $N$. The number of detectable single transits with $N \geq 10$ is found to be 336$\pm$26 (with two thirds coming from 30-minute cadence targets) and with $N \geq 100$ is 74$\pm$5 (with all but one target being at 2-minute cadence).

\subsubsection{2-minute cadence targets}
\label{sec:2-minute cadence targets}

Previous studies have found that when attempting to model orbital parameters from single transit events the sparsity of sampling within the transit is a big factor into the accuracy with which parameters can be obtained \citep{Osborn2015}. Because of this effect, transits sampled at 2-minute cadence will be better constrained. % Taking a simplified version of the equations presented in \citet{Osborn2015} (considering highest order effects) shows that $P \propto T_{dur}^3$. Taking this relation as a guide, and assuming that $\sigma_{T_{dur}}$ scales with cadence, it can be shown that $\sigma_P$ also scales with cadence. This means that \textit{TESS} candidates observed at 2-minute cadence may have periods constrained a factor of up to 15 better than those observed at 30-minute cadence. Therefore Table \ref{tab:period table2} shows the same information as Table \ref{tab:period table} but this time only planets sampled at a cadence of 2 minutes are included.
In the white-noise limit we can assume the following relation,

\begin{equation}
\label{eq:scaling}
\sigma_{T_{dur}} \propto \frac{1}{\sqrt{N}} \propto \Delta T.
\end{equation}

This means that \textit{TESS} candidates observed at 2-minute cadence may have periods constrained a factor of up to 15 better than those observed at 30-minute cadence. Therefore Table \ref{tab:period table2} shows the same information as Table \ref{tab:period table} but this time only planets sampled at a cadence of 2 minutes are included.
 
\begin{table}[ht]
\centering
\caption{Same information is shown as in Table \ref{tab:period table} except that only 2-minute cadence targets are included. Note that planets with $P<1$ day are omitted. Given values and uncertainties are the result of averaging multiple simulation runs.}
\label{tab:period table2}
\resizebox{\columnwidth}{!}{\begin{tabular}{|c|c|c|}
\hline
Period Bin (days) & Multitransits ($R_p \leq 4R_{\oplus}$) & Single Transits ($R_p \leq 4R_{\oplus}$) \\ \hline\hline
1.0 - 1.4         & 26$\pm$5   (24$\pm$4)        & 0$\pm$0  (0$\pm$0)         \\ \hline
1.4 - 2.0         & 47$\pm$8   (44$\pm$8)        & 0$\pm$0  (0$\pm$0)         \\ \hline
2.0 - 2.8         & 114$\pm$7  (101$\pm$7)       & 0$\pm$0  (0$\pm$0)         \\ \hline
2.8 - 4.0         & 115$\pm$17 (98$\pm$16)       & 0$\pm$0  (0$\pm$0)         \\ \hline
4.0 - 5.6         & 115$\pm$11 (92$\pm$8)        & 0$\pm$0  (0$\pm$0)         \\ \hline
5.6 - 7.9         & 210$\pm$12 (191$\pm$13)      & 0$\pm$0  (0$\pm$0)         \\ \hline
7.9 - 11.2        & 185$\pm$18 (164$\pm$19)      & 0$\pm$0  (0$\pm$0)         \\ \hline
11.2 - 15.8       & 167$\pm$24 (146$\pm$21)      & 3$\pm$1  (3$\pm$1)         \\ \hline
15.8 - 22.4       & 125$\pm$13 (110$\pm$10)      & 23$\pm$5 (20$\pm$4)        \\ \hline
22.4 - 31.6       & 73$\pm$9   (61$\pm$7)        & 38$\pm$4 (33$\pm$3)        \\ \hline
31.6 - 44.7       & 49$\pm$3   (39$\pm$4)        & 32$\pm$6 (24$\pm$5)        \\ \hline
44.7 - 63.1       & 12$\pm$4   (8$\pm$2)         & 10$\pm$5 (7$\pm$3)         \\ \hline
63.1 - 89.1       & 2$\pm$0    (2$\pm$1)         & 1$\pm$2  (1$\pm$1)         \\ \hline
89.1 - 125.9      & 1$\pm$1    (1$\pm$1)         & 1$\pm$1  (1$\pm$1)         \\ \hline
125.9 - 177.9     & 1$\pm$1    (0$\pm$0)         & 1$\pm$1  (0$\pm$0)         \\ \hline
177.9 - 251.2     & 0$\pm$0    (0$\pm$0)         & 0$\pm$0  (0$\pm$0)         \\ \hline
251.2 - 354.8     & 0$\pm$0    (0$\pm$0)         & 0$\pm$0  (0$\pm$0)         \\ \hline
354.8 - 501.2     & 0$\pm$0    (0$\pm$0)         & 0$\pm$0  (0$\pm$0)         \\ \hline
501.2 - 707.9     & 0$\pm$0    (0$\pm$0)         & 0$\pm$0  (0$\pm$0)         \\ \hline
707.9 - 1000.0    & 0$\pm$0    (0$\pm$0)         & 0$\pm$0  (0$\pm$0)         \\ \hline\hline
Total             & 1242$\pm$133 (1078$\pm$120)  & 133$\pm$24 (88$\pm$19)     \\ \hline
\end{tabular}}
\end{table}

Table \ref{tab:period table2} shows a decrease in the number of single transits as a fraction of all detectable planets when compared to both 2 and 30 minute cadence targets. All cadence targets show single transits to be $10.2\%$ of all detections whereas looking at only 2 minute targets shows single transits as $9.7\%$ of detections.

\subsection{Single site observations}
\label{sec:Single site observations}

\begin{figure*}[ht]
	\includegraphics[width=\linewidth]{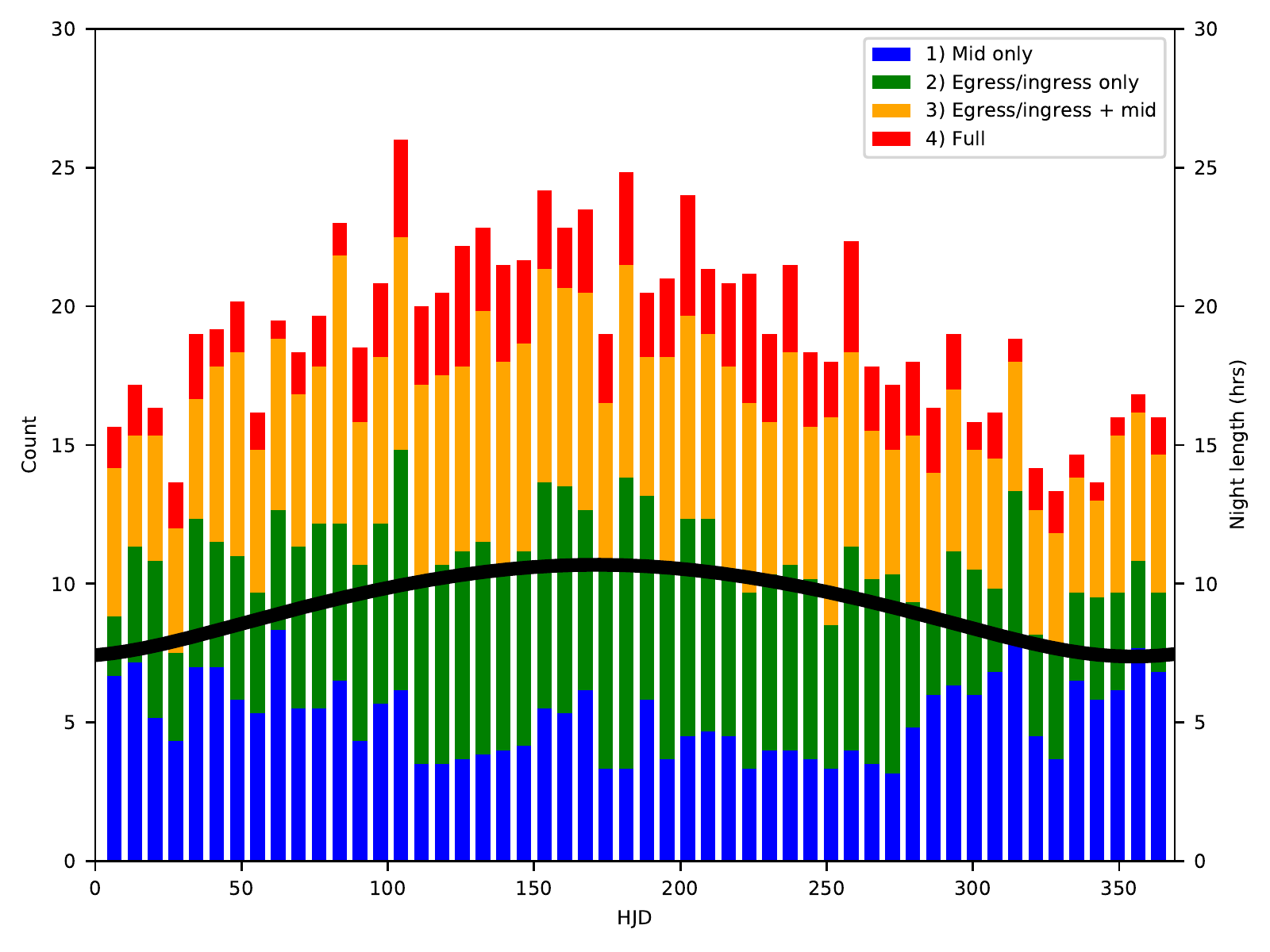}
    \caption{Distribution of \textit{TESS} single transits visible from Paranal for an arbitrary year. The data has been binned into weeks. The different colours correspond to different amount of the transit being visible. Colours blue, green, yellow and red corresponded to transits of the types 1, 2, 3 and 4 as described in the text. Additionally, the thick black line indicates the length of a Paranal night in hours. The HJD value is offset to run from 0-365 to cover 1 year of observations.}
    \label{fig:transit_dist}
\end{figure*}

When analysing the single transit detections from \textit{TESS} in terms of follow-up from a single site (here chosen to be Paranal) we find 1000$\pm$83 transit events observable in a single arbitrary year (320$\pm$21 unique targets) as seen in Fig. \ref{fig:transit_dist}. These events are broken down into the following categories due to how much of the transit occurs during a night. 
\begin{enumerate}
    \item 269$\pm$28 events where only the mid transit is observed (165$\pm$15 unique targets).
    \item 298$\pm$29 events where either an ingress or egress is seen (202$\pm$15 unique targets).
    \item 315$\pm$28 events when an ingress or egress is seen along with the mid transit point (201$\pm$19 unique targets).
    \item 117$\pm$12 events where the full transit is observed (83$\pm$10 unique targets).
\end{enumerate}
Figure \ref{fig:transit_dist} shows the distribution of transit events in an arbitrary year (binned into weeks) coloured by how much of the transit is observed. %Figure \ref{fig:transit_dist_sub} shows the same data split into 4 sub-plots for simplicity.

% \begin{figure*}[ht]
% 	\includegraphics[width=\linewidth]{figures/all_transits_subplots_noise2_redo_weekly.pdf}
%     \caption{Distribution of \textit{TESS} single transits visible from Paranal in an arbitrary year, binned into weeks. Each sub-plot shows a different subset of transits based on amount of the transit that is seen as described in the main text. Axes and colours are as in Fig. \ref{fig:transit_dist}. The length of a Paranal night in hours is shown in black in each sub-plot.}
%     \label{fig:transit_dist_sub}
% \end{figure*}

These events are spread through a year of observations (January to December) with noticeable changes towards the centre of the year coinciding with the increased length of a Paranal night. Of course, these observations do not yet take into account depth of the signal, magnitude of the host or moon illumination. The only requirement for an event to be listed here is that one transit is observable by \textit{TESS} and that a transit event is theoretically visible from Paranal during a night and at $\geq 30^{\circ}$ above the horizon within a single year of observations.

\begin{figure}[ht]
\centering
	\subfigure[\textit{TESS} detections deeper than 0.5\% transit depth. Shown here is one example simulation run.]{\label{fig:1}\includegraphics[width=\columnwidth]{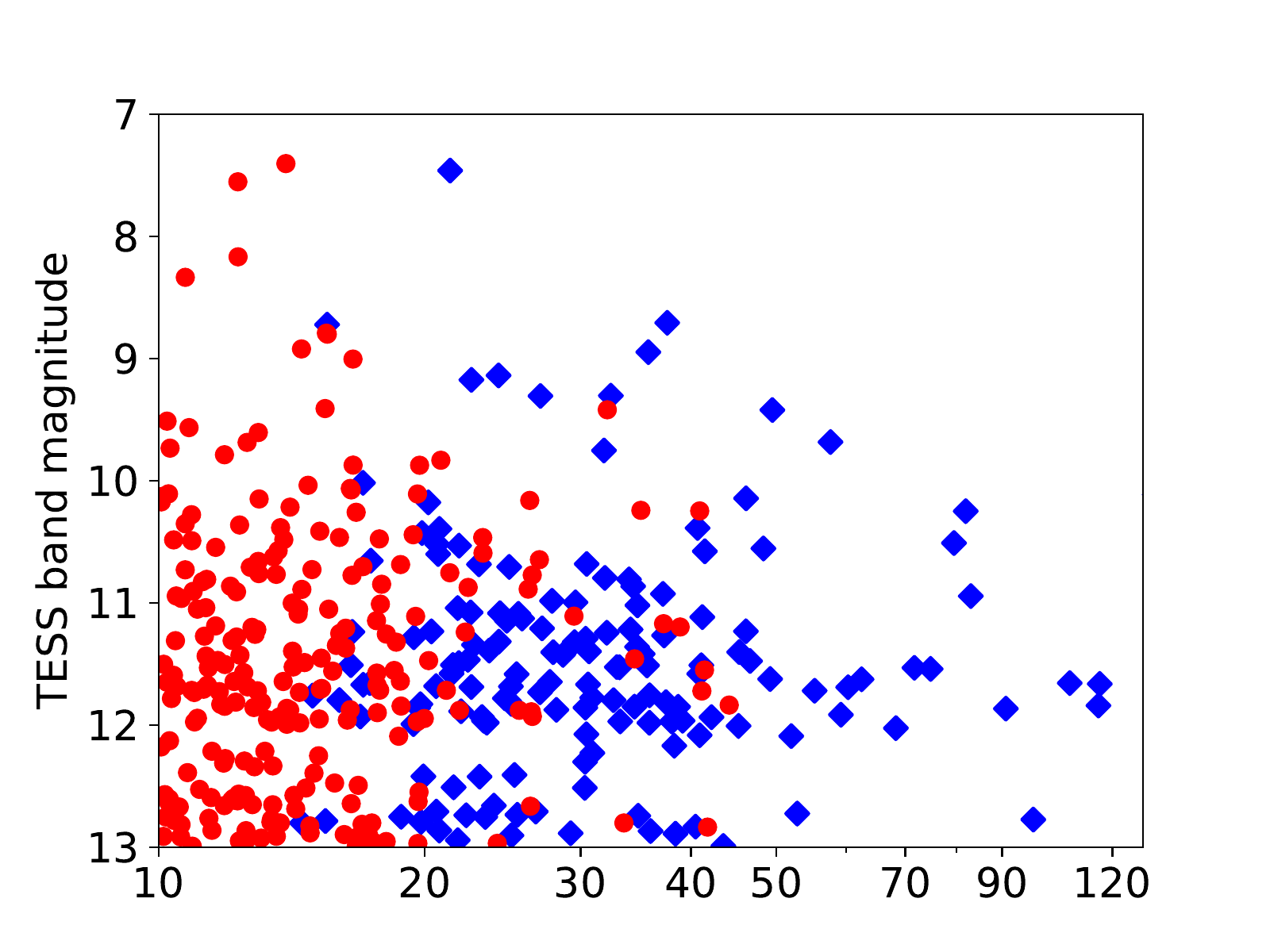}}
	\subfigure[\textit{TESS} detections histogram, averaged from multiple simulation runs]{\label{fig:2}\includegraphics[width=\columnwidth]{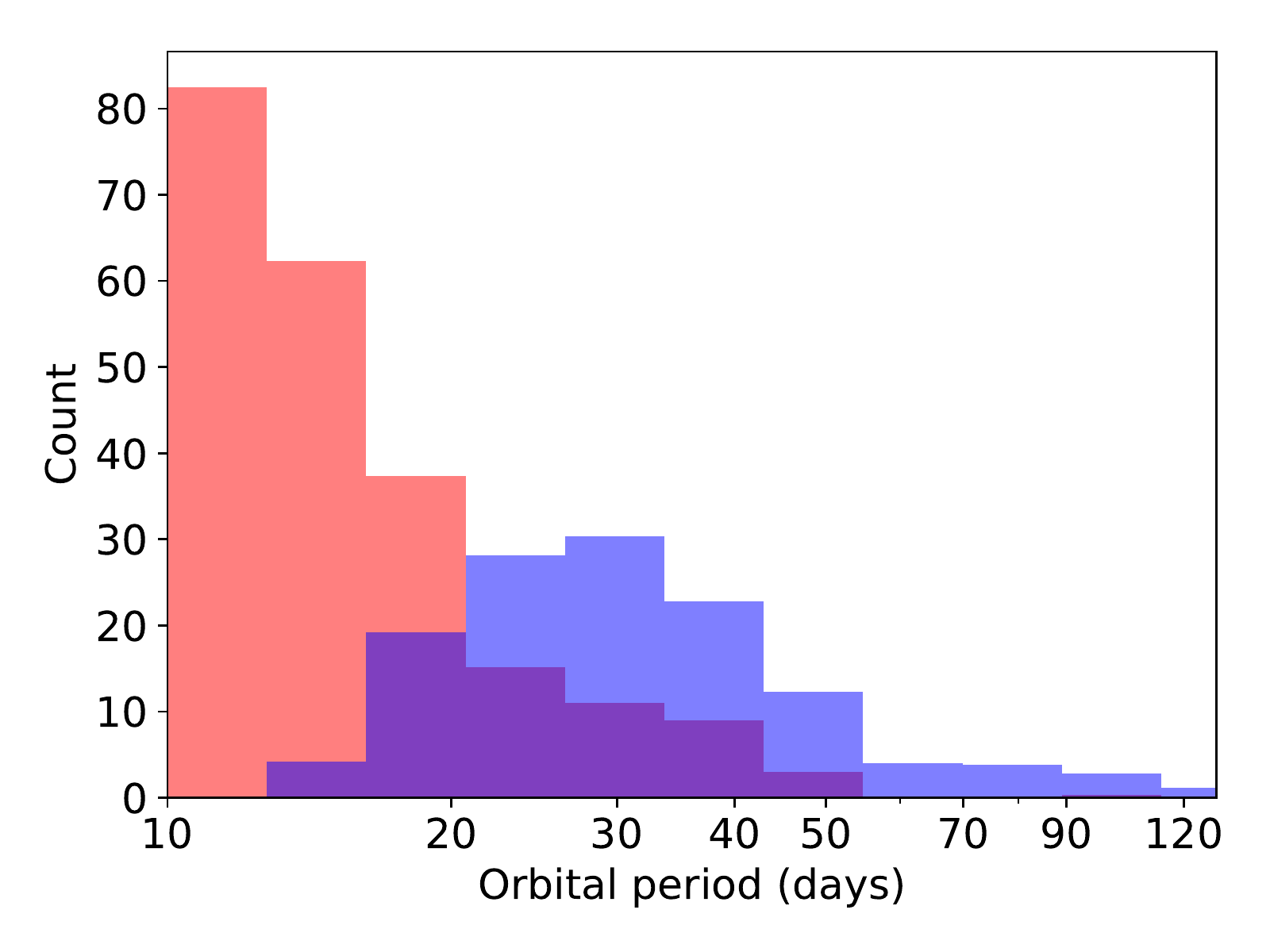}}
\caption{\textit{Top} All \textit{TESS} detections with at least 5\% transit depth. Additionally, this plot is restricted to $m_{TESS}\leq13$ and $P\geq10$ days. The red circles denote multitransit detections with blue diamonds showing single transits. \textit{Bottom} Corresponding histogram. Colours and limits are as before.}
\label{fig:pete_plot}
\end{figure}

Since the current follow-up observables are simply dependent on a single \textit{TESS} observation an additional cut was made to focus on signals that are realistically observable from ground-based sites. In the advent of the latest ground-based transit surveys such as \textit{NGTS}, sub-percent transit depths are possible from single transits. Therefore, a cut was made at $\delta_{eff}\geq 0.005$, or a 0.5\% transit depth. The remaining candidates are shown in Fig. \ref{fig:pete_plot}.

As can be seen from Fig. \ref{fig:pete_plot}, there are still a significant number of detections even when only looking at relatively deep transits. Also key to note is that at long periods ($P\geq40$ days) there are no multitransit detections (0$\pm$0) but a significant number (30$\pm$5) of single transits. This confirms that study of this subset of exoplanets will require follow-up of \textit{TESS} single transit detections. Additionally, even at shorter period ranges ($20\leq P<40$ days) the brightest detections will come from single transits. Of the brightest ($m_{TESS}\leq 11$) detections in this period range 18$\pm$3 out of 18$\pm$3 will come from single transits (values determined by averaging across multiple simulation runs). Bright targets are of particular interest as they will be most amenable to atmospheric characterisation via transmission spectroscopy with \textit{JWST}.

In the work laid out so far we have assumed perfect knowledge of candidate period from a single \textit{TESS} transit observation. In the following section we relax this assumption and consider realistic ephemeris uncertainties. We also now consider only the best \textit{TESS} single transit detections, that is, those targets with a transit depth of at least 0.5\% and a stellar magnitude less than 13 in \textit{TESS} band.

\subsection{Realistic follow-up}
\label{sec:Realistic follow-up}

Since the targets we now consider have relatively deep transits and are hosted by relatively bright stars we assume that the uncertainty on this value is likely to be approximately 10\% \citep{Osborn2015} (this value applies to favourable single transits but here we only consider the best subset). Based on this value we can make some predictions regarding the time required to follow-up a single transit detection. To effectively follow-up a candidate we assume that it is necessary to capture the entire duration of the transit event. To calculate this time allowing for uncertainties in measured parameters we first assume a flat error of 10\% on period. Next we make the assumption that the follow-up efforts are carried out on the next transit of a given planet (a challenge, but feasible, especially for the longer period targets). Therefore, assuming the ephemeris of the \textit{TESS} detection is known to sufficient accuracy, the total time required to observe a single candidate effectively is equal to its transit duration plus a buffer period at either end of the predicted ephemeris to account for the 10\% uncertainty in period. Thus the total time required to effectively follow-up a candidate, $T_{total}$, is given by

\begin{equation}
\label{eq:Ttotal}
T_{total} = 2\left(\frac{P}{10}\right) + T_{dur}.
\end{equation}

\begin{figure}[ht]
	\includegraphics[width=\columnwidth]{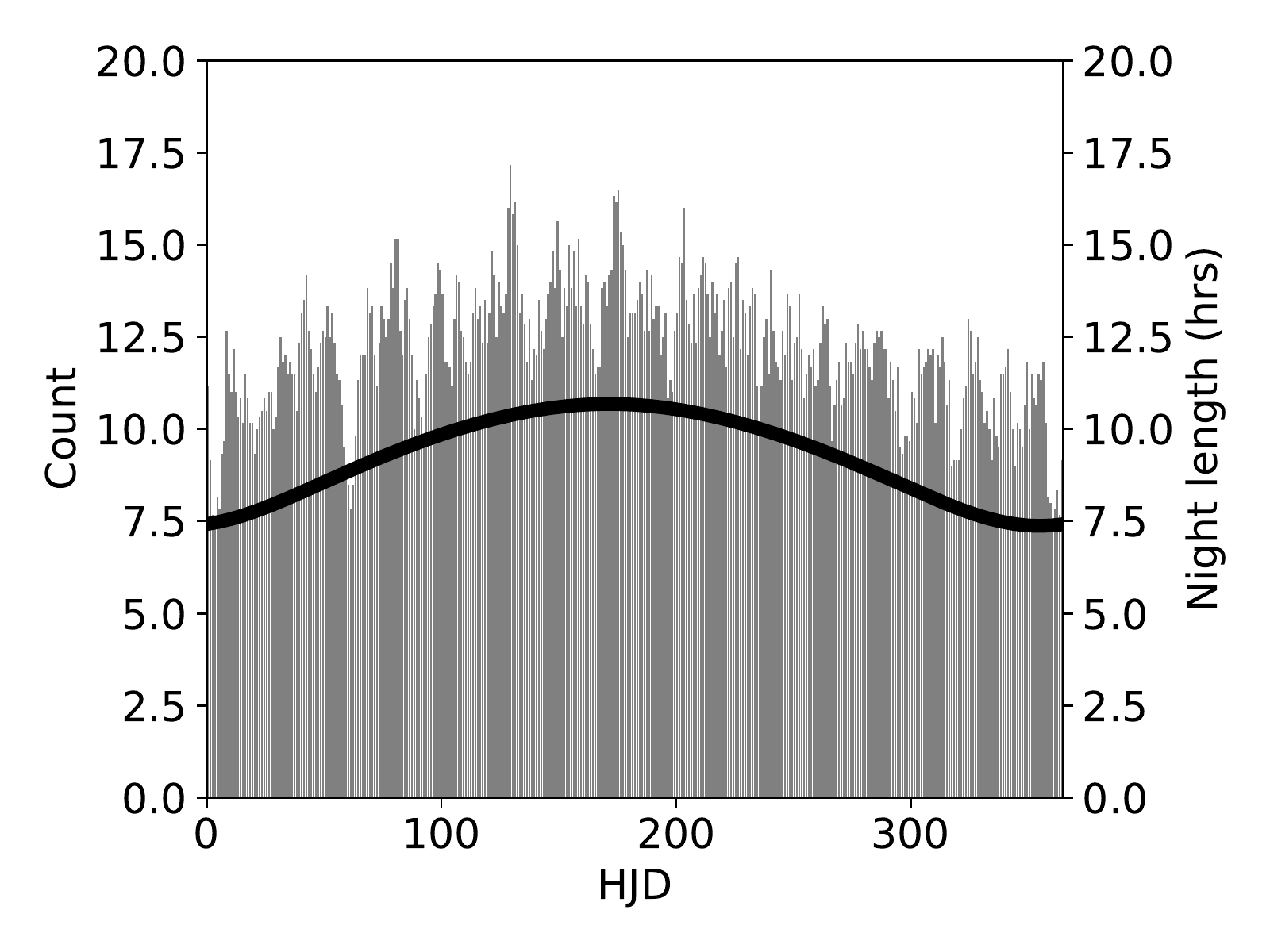}
    \caption{Distribution of the best \textit{TESS} single transits visible from Paranal for an arbitrary year. Additionally, the thick black line indicates the length of a Paranal night in hours. The HJD value is offset to run from 0-365 to cover 1 year of observations. The values here are the average of multiple simulation runs.}
    \label{fig:transit_dist_review}
\end{figure}

Calculating this value for the best \textit{TESS} single transit detections averaged over multiple simulation runs (130$\pm$13 targets) gives a total of 22498$\pm$2867 hours or an average of 173$\pm$9 hours per candidate. This is a significant amount of time but it can easily be reduced via a few simple measures. For example, since the majority of this time is caused by the uncertainty in period, large amounts of observation time will be out of transit and show no variation. Thus, multiple targets can be observed concurrently using an on-off technique, switching between them until one begins to transit at which point that target is observed continuously. Figure \ref{fig:transit_dist_review} shows a plot of the distribution of observable transit events from Paranal in a year as in Fig. \ref{fig:transit_dist}. This plot however, considers only the best \textit{TESS} candidates as described above and uses the the total time required for follow-up (Eq. \ref{eq:Ttotal}) instead of the exact transit duration.

From figure \ref{fig:transit_dist_review} it can be seen that there are many more events (due to the increased effective transit length). In fact, there are never fewer than 7 events per night and up to 17 in some nights with an average of 12.  The large number of simultaneous events means that the on-off technique described above will be able to monitor multiple transit events concurrently using a single telescope, reducing the total telescope time needed.

Additionally, we may utilise existing archival data (such as the publicly available photometry published by \textit{SuperWASP} and \textit{NGTS}) as shown by \citet{Yao2018}. Using this technique may drastically improve the ephemeris and then allow for a more accurate prediction of subsequent transits. The best targets we have selected here have the added bonus that they will most likely produce signals observable in archival data meaning they are the most likely targets to receive reduced ephemeris uncertainties via this method. Finally, this time may be further reduced via the use of spectroscopic follow-up, although it should be noted that spectroscopy alone will not provide accurate period estimates, only allow for better timing of ground-based photometry efforts.

Beyond selecting Paranal as the single site this paper does not go into any specifics about the potential observatory that may be used for these follow-up efforts. Obviously the observatory will have an impact on the observations due to factors including field of view, photometric precision and availability of telescope time. Not discussing these details here is a deliberate choice to keep the results of this paper non-specific. Therefore, these results can then be applied to any observatory and refined as necessary.

\subsection{False positives}
\label{sec:False positives}

In this paper we do not conduct any analysis of false positive single transit detections in the \textit{TESS} data. For a proper analysis of sources and rates of false positives see \citet{Sullivan2015}. The results of \citet{Sullivan2015} imply a number of false positives comparable to the number of genuine detections. The choice of S/N value used in this paper is chosen as an antidote to these spurious detections. This threshold is designed to result in 1 statistical false positive per 2$\times$10$^5$ light curves when looking at multiply transiting planets \citep{Bouma2017}. For single transit detections it will be harder to separate false positives, therefore Fig. \ref{fig:d} is of particular use, allowing us to see how increasing the required threshold will affect our single transit yields. Additionally, the best targets considered in Fig. \ref{fig:pete_plot} are all relatively deep, further reducing the risk of them being false positives. Since the time required to follow-up our candidates is significant it is important to be able to separate real signals from false so as to avoid wasted time. However, the results of this simulation show an appreciable number of detections, such that it should be feasible to select only the most convincing signals and still maintain a sufficiently large sample. Finally, Fig. \ref{fig:transit_dist_review} shows that multiple targets may be observed using the on-off method described in Sec. \ref{sec:Realistic follow-up}, further reducing the impact of false detections.

\section{Discussion and conclusions}
\label{sec:Discussion and conclusions}

The TIC CTL contains approximately 3.8 million stars, after making two cuts for $m_{TESS}$ and $T_{eff}$ approximately 3.6 million stars remain. We simulated approximately 4 million stars orbiting the 3.8 million TIC CTL stars. Of these planets more than 4700 will have detectable transits during \textit{TESS} observations and around 460 will transit only once.

%As a display of the discovery potential of \textit{TESS}, Fig. \ref{fig:NASA_planets} shows how the predicted \textit{TESS} distribution of planets compares to those already confirmed.

%\begin{figure}[ht]
%	\includegraphics[width=\columnwidth]{figures/NASA_planets_noise2_redo}
%    \caption{Plot of orbital separation against orbital period for predicted \textit{TESS} discoveries and known exoplanets. The \textit{TESS} planets are separated into multiple and single transits and the confirmed planets are sourced from the list of known transiting planets in the NASA exoplanet archive.}
%    \label{fig:NASA_planets}
%\end{figure}

We find that 320$\pm$21 planets will have a least one transit event observable from Paranal within a 1 year period. As discussed in Sect. \ref{sec:Single site observations} it can be seen that there are an average of $\sim3.0$ transit events per night throughout 1 year (however, this rate is not uniform, and peaks around the centre of the year). It is found that the number of full transits seen is $\sim0.3$ per night (this value also peaks towards the centre of the year).

However, it is also important to note that the numbers presented in this report are all lower bounds. The simulation presented above only simulates planets around known CTL stars, however each full frame image will contain stars absent from the CTL, any one of which could also display a transit during observations, thus the expected number of transits, and therefore single transits, is higher than those presented here. The CTL stars have been chosen as the best candidate hosts from the TIC \citep{Stassun2017} but it is possible for non-CTL stars to exhibit observable, and detectable, transits. As a comparison, \citet{Sullivan2015} conduct a \textit{TESS} detection simulation using a simulated population of stars rather than a known catalogue. From this data they expect over 20,000 detectable \textit{TESS} transits. Assuming single transits scale as they do in this paper ($\sim10\%$ of detections) the actual number of single transits found by \textit{TESS} may be closer to 2000. Additionally, this study neglects secondary transits. Obviously these are much smaller and harder to detect than primary transits but have the effect of doubling the potential events in the same time period. It is possible that the best targets may not exhibit a primary transit during observations but may host a planet whose secondary transit is observable at the photometric precision of \textit{TESS}.

As well as the raw numbers of single transits that will be found with \textit{TESS} it is important to consider these numbers in comparison to the multitransits found. single transits are harder to follow-up as it is more difficult to obtain accurate transit parameters based on a single transit, although it has been shown that accurate predictions can be made, especially if circular orbits are assumed \citep{Osborn2015}. With this in mind the most important result of this simulation is that presented in Table \ref{tab:period table} where multiple and single transit detections are compared in period bins. From this table it can be seen that for period of $P\gtrsim30$ days the number of single transits outweighs the multitransit detections, though the number of multitransit detections is still significant. For periods $P\gtrsim45$ days however the numbers begin to seem more favourable to single transit follow-up. In fact, by the time we reach $P\gtrsim60$ days \textit{TESS} is predicted to find only 6$\pm$3 multitransit detections but 28$\pm$11 single transit detections. The difference between these two numbers is much more significant and is a strong argument that following up these single transit detections will afford opportunities to explore science that would not be possible if only multitransit \textit{TESS} detections are followed-up.

One possible avenue of science that may be better explored is the discussion of long period ($P\gtrsim45$ days) sub-Neptune/super-Earth planets. Recent results have repeatedly shown a dearth of planets around $2R_{\oplus}$ \citep{Lundkvist2016,Fulton2017,VanEylen2017}, potentially a consequence of photoevaporation \citep{Owen2013,Lopez2013,Owen2017}. Therefore, \textit{TESS} detections of planets within this region may be able to aid in this discussion since they will likely be detected around relatively bright hosts. To this end, Table \ref{tab:period table} includes an additional break down showing numbers of simulated \textit{TESS} detections which are predicted to be found with $R_p \leq 4R_{\oplus}$. When looking at this particular period and radius range \textit{TESS} multitransits will yield  13$\pm$4 targets and \textit{TESS} single transits will yield 14$\pm$6. The increase in a statistical sample from 13 to 27 is significant and is a good reason to spend additional effort following up \textit{TESS} single transits. It should also be noted that these detections are doubly important as, since they are around \textit{TESS} candidate hosts, they are brighter than average which will lead to improved prospects for follow-up characterisation including atmospheric identification with future missions \citep{Louie2018,Batalha2017}.

\subsection{Comparison with previous work}
\label{sec:Comparison with previous work}

The numbers presented in this paper are in excellent agreement with those presented in \citet{Barclay2018}. We find an approximate 10\% increase in \textit{TESS} detections which is as expected as these additional detections come from the extension of the simulation up to longer periods and the inclusion of single transiting planets as candidates. When compared with another recent estimation of \textit{TESS} single transits, \citet{Villanueva2018}, we find fewer detections. However, this is most likely due to the different approaches used. Whereas this paper employs a Monte Carlo simulation (similar to the method employed by \citealt{Barclay2018}), \citet{Villanueva2018} use a more analytic integration based approximation. In said paper the authors acknowledge that they may over estimate detections compared to a more realistic simulation as employed in \citet{Barclay2018} (and thus, in this paper) therefore we do not find the discrepancies between results to be an issue.

%\begin{acknowledgements}
%      
%\end{acknowledgements}

%-------------------------------------------------------------------
\bibliographystyle{aa}
\bibliography{TESS}

\end{document}